\documentclass[twocolumn]{aastex631} %linenumbers for submission

\usepackage{amsmath}
\shorttitle{AASTeX v6.3.1 Sample article}
\shortauthors{Cavieres et al.}
\graphicspath{{./}{}}
\begin{document}

%\title{Characterization of the LMC induced dynamical friction Wake}
\title{The distant Milky Way halo from the Southern hemisphere: Characterization of the LMC-induced dynamical-friction wake\footnote{Based on observations collected at NOIRLab Cerro Tololo Inter-American Observatory under Chilean TAC program CN2020B-18, and at the European Southern Observatory under ESO programme 109.23H5.001}}

\correspondingauthor{Manuel Cavieres}
\email{mncavieres@uc.cl}

\author[0009-0002-2978-8383]{Manuel Cavieres}
\affiliation{Instituto de Astrof\'{i}sica, Pontificia Universidad Cat\'{o}lica de Chile, Av. Vicu\~{n}a Mackenna 4860, 782-0436 Macul, Santiago, Chile}
%\affiliation{Leiden Observatory, Leiden University, P.O. Box 9513, NL-2300 RA Leiden, the Netherlands} % Esto en verdad depended de mi contrato y lo estoy revisando

\author[0000-0003-2481-4546]{Julio Chanam\'e}
\affiliation{Instituto de Astrof\'{i}sica, Pontificia Universidad Cat\'{o}lica de Chile, Av. Vicu\~{n}a Mackenna 4860, 782-0436 Macul, Santiago, Chile}

\author[0000-0002-4777-9934]{Camila Navarrete}
\affiliation{Observatoire de la C\^{o}te d'Azur,
Boulevard de l'Observatoire, 06304 Nice, France}

\author[0000-0001-7966-7606]{Yasna Ordenes-Brice\~{n}o}
\affiliation{Instituto de Estudios Astrof\'{i}sicos, Facultad de Ingenier\'{i}a y Ciencias, Universidad Diego Portales, Av. Ej\'{e}rcito Libertador 441,
Santiago, Chile}

\author[0000-0001-7107-1744]{Nicol\'as Garavito-Camargo}
\affiliation{Center for Computational Astrophysics, Flatiron Institute, 162 5th Ave, New York, NY 10010, USA}

\author[0000-0003-0715-2173]{Gurtina Besla}
\affiliation{University of Arizona, 933 N. Cherry Ave, Tucson, AZ 85721, USA}

\author[0000-0002-2982-8712]{Maren Hempel}
\affiliation{Universidad Andr\'es Bello, Fernandez Concha 700, Las Condes, Santiago, Chile}
\affiliation{Max-Planck Institute for Astrophysics,Königstuhl 17, 69117 Heidelberg, Germany .}

\author[0000-0003-4341-6172]{A. Katherina Vivas}
\affiliation{Cerro Tololo Inter-American Observatory/NSF’s NOIRLab, Casilla 603, La Serena, Chile}

\author[0000-0002-1947-333X]{Facundo G\'omez}
\affiliation{Universidad de la Serena, Avenida Cisternas 1200, La Serena, Chile}



\begin{abstract}

The infall of the Large Magellanic Cloud (LMC) into the Milky Way's halo impacts the distribution of stars and dark matter in our Galaxy. Mapping the observational consequences of this encounter can inform us about the properties of both galaxies, details of their interaction, and possibly distinguish between different dark matter models. N-body simulations predict a localized overdensity trailing the LMC's orbit both in baryonic and dark matter, known as the wake. We collected wide-field, deep near-infrared, and optical photometry using VIRCAM and DECam across four fields along the expected wake, covering the sky region expected to span most of its predicted density contrast. We identify over 400 stars comprising two different tracers—near main sequence turn-off stars and red giants—that map the halo between ~60–100 kpc, deriving stellar halo densities as a function of sky position and Galactocentric radius. We detect (1) a break in the halo radial density profile at 70 kpc not seen in Northern halo studies, and (2) a clear halo overdensity starting also at 70 kpc, with density contrast increasing steadily toward the expected current location of the wake. If this overdensity is the LMC wake, its peak density contrast is as pronounced as the most massive LMC model considered. Contamination from unidentified substructures may bias our wake detections, so wider-area surveys with similar depth are needed for confirmation.

\end{abstract}



\section{Introduction} \label{sec:intro}

With the advent of space-based proper motion measurements of the LMC/SMC system \citep{2006ApJ...638..772K,2006ApJ...652.1213K}, it has been found that they are most likely on their first passage about the Milky Way \citep[MW,][]{2007ApJ...668..949B}, although the possibility of a secondary passage may hold still  \citep{2024MNRAS.527..437V}. However, as shown by \citep{2023Galax..11..114Z} a second passage can not explain the latitudinal velocity of RR Lyrae. In a first infall scenario, the impact of the LMC on the MW dark matter (DM) halo will be relatively recent ($<$1 Gyr), allowing us to probe the imprint of its passage in the stellar halo and potentially test Cold Dark Matter (CDM) theory.

In particular, the LMC is the MW's most massive satellite, with current mass estimates placing it at about 10\% of the MW mass \citep{2010ApJ...721L..97B, 2016MNRAS.456L..54P, 2017MNRAS.464.3825P, 2019MNRAS.487.2685E, 2020MNRAS.495.2554E, 2021ApJ...923..149S, 2021MNRAS.501.2279V, 2022MNRAS.511.2610C, 2023MNRAS.521.4936K}. Since the seminal work of Chandrasekar \citep{1943ApJ....97..255C}, standard theory predicts that the passage of a massive object will create a density wake due to the direct gravitational scattering of the particles in the medium, which in turn acts on the massive body itself, decelerating it by a transfer of angular momentum. This phenomenon, also referred to as dynamical friction (DF), when applied to the passage of the LMC through the MW halo involves both a dark matter wake and a stellar wake counterpart. In this article, we will refer to the stellar wake induced by the passage of the LMC on the MW halo as the \textit{wake}. 

The infall of the LMC towards the MW yields several significant effects. Such as: radial and on-sky angular dependant perturbations to MW's streams \citep{ 2019MNRAS.485.4726K, 2019MNRAS.487.2685E, 2019ApJ...885....3S, 2021ApJ...923..149S, 2024arXiv241002574B}; the MW's outer halo reflex motion with respect to the MW--LMC's barycenter \citep{2015ApJ...802..128G, 2021NatAs...5..251P, 2021MNRAS.506.2677E, 2024MNRAS.531.3524Y, 2024arXiv240601676C}; a collective response characterized by a large-scale overdensity leading the LMC, which primarily results from the displacement between the inner and outer halo;  a global underdensity that envelops most of the southern hemisphere, encircling the dynamical friction wake, alongside the DF wake overdensity itself \citep{2019ApJ...884...51G, 2021ApJ...919..109G, 2021ApJ...916...55T, 2022ApJ...933..113R}. 

Observational evidence of the presence of the wake has been found in the so-called "Pisces overdensity" (PO), first identified as an overdense region in SDSS strip 82 \citep{2007AJ....134.2236S}, corresponding to the largest overdense region (in volume) after the Sagittarius Stream. Both RR Lyrae and blue horizontal branch (BHB) stars have been observed in the Pisces overdensity \citep{2015ApJ...810..153N, 2009MNRAS.398.1757W,2018ApJ...862L...1D, 2019MNRAS.488L..47B} tracing an elongated shape and distance gradient ranging from 40 kpc up to 100 kpc. Radial velocities of 13 BHB stars with typical halo metallicity in the Pisces Overdensity are consistent with those predicted in \citet{2019ApJ...884...51G}, suggesting an association with the stellar wake induced by the LMC. The structure formed by stars at the edge of the PO region, consistent with the LMC stellar wake, has been dubbed the Pisces Plume and has been found to extend from 40 to 100 kpc. \citep{2019MNRAS.488L..47B}. 

%Additionally, \cite{2021Natur.592..534C} reported the detection of the local wake and the collective response using photometry from Gaia Early Data Release 3 \citep{2021A&A...649A...5F, 2021A&A...649A...3R}, and infrared photometry from the Wide-field Infrared Survey Explorer \citep[WISE,][]{2010AJ....140.1868W}. Their study comprises an analysis of 1301 photometrically selected K giants ($\sim 0.03\,{\rm stars/deg}^2$) to produce an all-sky density map, revealing an over-density with an apparent amplitude exceeding predictions by numerical models \citep{2019ApJ...884...51G}.  Recent work by \citet{amarante24} has identified a large sample of BHB stars reaching distances of up to 120 kpc and reported a signature of the wake caused by the LMC. They measured a density contrast of $\geq$0.6 across a region approximately 60 degrees long and 25 degrees wide, aligned with the LMC's orbit.
Additionally, \citet{2021Natur.592..534C} reported the detection of the local wake and the collective response using photometry from Gaia Early Data Release 3 \citep{2021A&A...649A...5F, 2021A&A...649A...3R} and the Wide-field Infrared Survey Explorer \citep[WISE;][]{2010AJ....140.1868W}. Their analysis revealed an over-density in an all-sky density map, with an amplitude exceeding predictions from numerical models \citep{2019ApJ...884...51G}. Recent work by \citet{amarante24}, using Legacy Survey DR9 photometry \citep{2019AJ....157..168D}, identified a large sample of BHB stars reaching distances of up to 120 kpc and reported a signature of the wake caused by the LMC. They measured a density contrast of $\geq 0.6$ across a region approximately 60 degrees long and 25 degrees wide, aligned with the LMC's orbit.

%Furthermore, \citet{2023ApJ...956..110C} argue that the over-density reported by \cite{2021Natur.592..534C} is affected by debris from the LMC and SMC, with at least 20\% and up to 50\% of the objects being associated with the Clouds. Therefore, an independent detection using more abundant stellar tracers becomes necessary. This will provide a larger, statistically significant trace of the wake and its possible density enhancements along the MW halo, and along the line of sight.

Moreover, the characteristics and density structure of DF wakes are influenced by the physical properties of the medium in which they develop \citep{1999ApJ...525C.297O, 2002ApJ...565..854F, 2020JCAP...01..001L, 2023JCAP...04..071V}. Consequently, the wake serves as an ideal laboratory for studying the microphysics of DM \citep[see][for a comparison of the wake in fuzzy DM and cold DM]{Foote_2023}. The density and structure of the wake are dependent on both the mass of the LMC and MW, as well as the orbit of the LMC.  

In this paper, we report the results of our program for detecting and characterizing the wake through an observational campaign combining both near-infrared and optical photometry. Our data allows us to provide a detection of the wake independent of data and results from previous detections, as well as a detailed view of the specific density enhancement at different heliocentric distances. Based on the observed density enhancements, we compare with existing simulations of the LMC-MW interaction, which allows us to place constraints on the virial mass at infall of the LMC. 

This paper is structured as follows: in Section \ref{sec:obs} we describe the observations, data reduction procedures, and photometry methodologies. Section \ref{sec:tracers} describes the selection criteria applied to identify the halo tracers used, as well as the methodology to derive photometric distances; the details of the simulations used in this work are described in Section \ref{sec:sim}; In Section \ref{sec:results} we present our derived halo density profiles and their comparison with simulations. The discussion takes place in Section \ref{sec:discussion}. We conclude in Section \ref{sec:conclusions}.

\section{Observations and data processing}\label{sec:obs}

\begin{figure*}
    \centering
    \includegraphics[width = 0.7\linewidth]{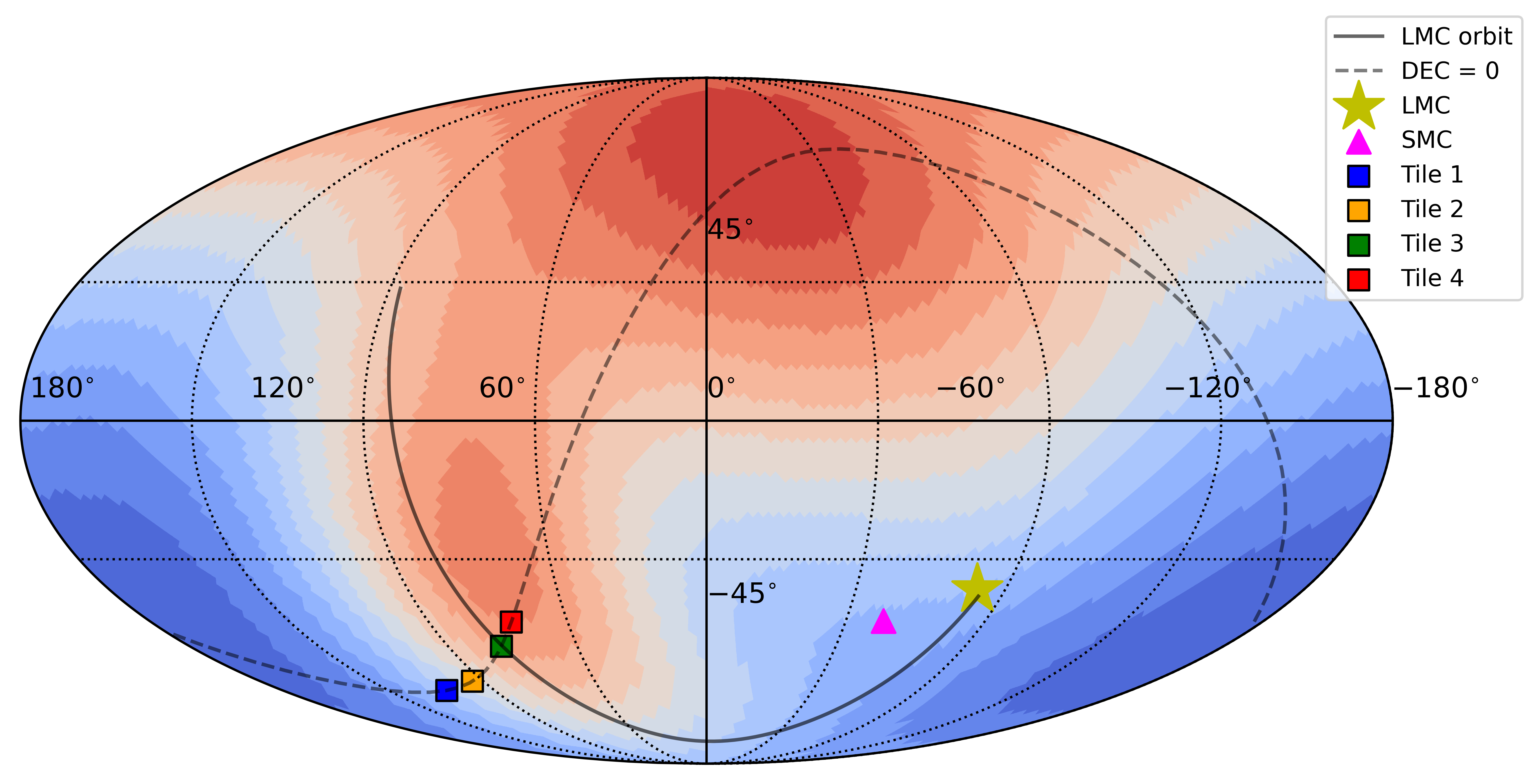}
    %\label{fig:observations_plot}

    \vspace{0.3cm}
    \includegraphics[width = 0.7\linewidth]{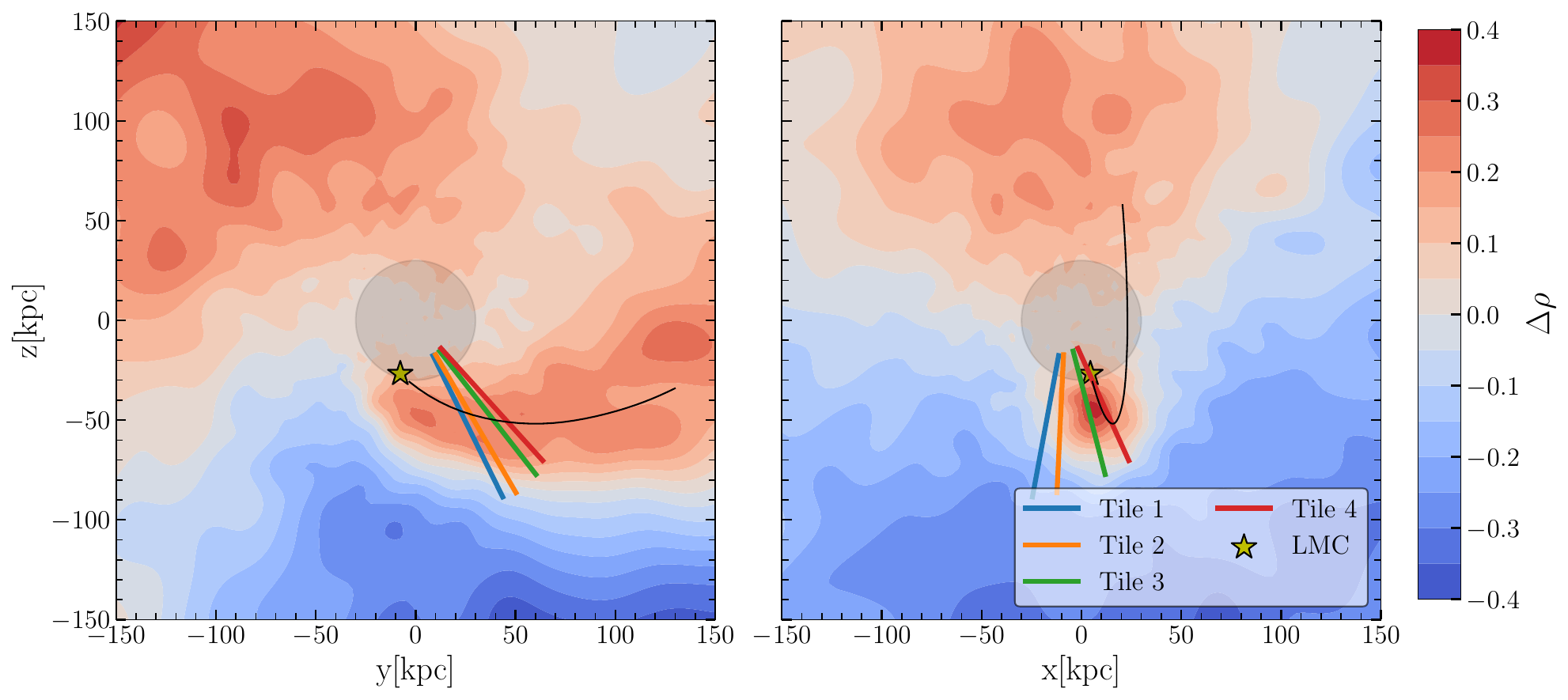}
    %\label{fig:projections_tiles}
    
    \caption{Upper panel shows the all-sky Aitoff projection map of simulation particles (at Galactocentric distances in the range 60 kpc $< R_{\rm GC} <$ 100 kpc) from \citet{2019ApJ...884...51G} models of the LMC-MW interaction. Displayed in blue-to-red is the dark matter particle density for a \citet{2019ApJ...884...51G} simulation that includes the interaction between the LMC and MW-halo. Density is computed as the number of particles found per square degree based on the fifth-level HEALPix tesselation system \citep{healpix} smoothed with a 15-degree kernel. The lower panels show the projected DM density in the Galactocentric plane at x = 0 on the left and y = 0 on the right, both with a thickness of 10 kpc. The overdensity in the north is the collective response of the Galactic halo, while the overdensity found in the southeast region corresponds to the wake. Our observed fields are shown as squares and lines in blue, yellow, green, and red, corresponding to Tile 1, 2, 3, and 4. The SMC and LMC are shown as a magenta triangle and yellow star, respectively, for reference purposes. The past orbit of the LMC obtained from \citet{2019ApJ...884...51G} is shown as a black line, and the celestial equator is shown as a dotted grey line. It can be clearly observed that, when comparing the density across our observed fields, we expect to find a density enhancement both along the line of sight, as shown in the lower panels, and in projection, as shown in the upper panel. }
    \label{fig:observations_plot}
\end{figure*}

The primary observational data used in this paper were collected as part of a photometric campaign carried out during 2020 at the CTIO 4m Blanco telescope with the Dark Energy Camera \citep[DECam]{2015AJ....150..150F} along with data collected using the Visible and Infrared Survey Telescope for Astronomy \citep[VISTA;][]{2006Msngr.126...41E} and the VIRCAM instrument \citep{2006SPIE.6269E..0XD}. We observed 4 fields to which we will refer by their tile number from 1 to 4. The pointing coordinates are at DEC = 0 deg, RA = $334.9$,  $343.0$, $359.0$, $7.0$ deg, and were defined to approximately cover the entire density range of the wake according to current numerical models of the LMC-MW interaction \citep{2019ApJ...884...51G} and also probing the overdensity known as the Pisces Overdensity, and delving into the Pisces Plume in tile 4 \citep{2019MNRAS.488L..47B, 2021Natur.592..534C}. Figure \ref{fig:observations_plot} presents the locations of the 4 observed fields overlayed on a density map of DM particles from \cite{2019ApJ...884...51G}. For convenience, the location of the observed fields was placed along the celestial equator (dotted grey line in Figure \ref{fig:observations_plot}) as it is approximately perpendicular to the expected past orbit of the LMC shown as a black line in Figure \ref{fig:observations_plot}. 

DECam observations were conducted from November 9-12, 2020, with additional exposures on November 23, 2020. The strategy aimed to maximize the signal-to-noise ratio for faint sources while minimizing overheads. All \textit{g} and \textit{i} band exposures were taken during dark time, with 3600 seconds for \textit{g} and 4800 seconds for \textit{i}, split into 10-minute intervals, totaling 10 hours of integration (2.5 hours per field). Seeing ranged from $0".58$ to $1".04$, with medians of $0".68$ in \textit{g} and $0".94$ in \textit{i}. The airmass varied from 1.0 to 1.59, with medians of 1.24 in \textit{g} and 1.15 in \textit{i}. A 5-arcsecond line-dither pattern was used, improving bad pixel interpolation but not covering gaps between detectors, resulting in a larger effective exposure for stacked images.

Near-infrared VIRCAM observations took place from June 2 to November 25, 2022, in the $K_s$ band, with 7-second integration times per exposure in a 12-exposure `jitter3u' pattern, totaling 84 seconds per \textit{pawprint}. Using the standard 16-pawprint pattern, we covered $1.6 {\rm deg}^2$ per field, matching the DECam field of view. In total, 2516 pawprints (58.7 hours) were collected, with about 14.7 hours per field. The airmass ranged from 1.1 to 1.723 (median 1.22), and seeing varied between $0".57$ and $0".99$, with a median of $0".8$.

\subsection{Image processing}

The DECam data was processed using the NOAO Community Pipeline \citep{2014ASPC..485..379V}, which produced the stacked images. Due to long exposure times, many observations were affected by satellite trails. To address this, a 15-pixel median blur was applied to reduce noise, followed by a Canny filter \citep{4767851} for edge detection. The edge-detected image was converted to binary using an adaptive Gaussian threshold, based on the 11x11 pixel neighborhood around each pixel, manually adjusted by subtracting 2. A probabilistic Hough transform \citep{hough} identified lines (minimum 100 pixels, maximum 50-pixel gaps), and all identified lines were masked in the weight maps. For VIRCAM, pawprints with satellite stripes were excluded.

VIRCAM data analysis relied on the standard data products from the VISTA Data Flow System \citep[VDFS;][]{2004SPIE.5493..401E}, processed by the Cambridge Astronomy Survey Unit (CASU) pipeline \citep{2004SPIE.5493..411I}. We resampled processed pawprints using \textit{swarp} \citep{swarp} with a LANCOZ3 kernel.

Pawprint coaddition was performed with \textit{swarp} \citep{swarp}, using a clipped combination algorithm. This algorithm detects outliers, applies a mask, and then performs a median combination to reduce ghosting and cosmic ray artifacts.

\subsection{Photometry} \label{sec:phot}
Photometry was performed using \textit{SExtractor} and \textit{PSFex} \citep{sextractor, psfex}. The process began with image segmentation to set an accurate detection threshold, focusing on the detector-covered areas of the stacked image. This step is crucial as gaps between chips can skew standard deviation estimates from \textit{SExtractor}. Only pixels exceeding the \texttt{SKYSUB} threshold in the image header were considered. The standard deviation was then calculated after five iterations of sigma clipping using a $5 \sigma$ limit to exclude obvious sources.

Source detection followed, identifying sources above 1.5 standard deviations across more than 3 pixels. Their positions, ellipticity, FWHM, and fluxes were measured. Sources with \textit{SExtractor} $\texttt|FLAG| = 0$, $\texttt{SNR} > 20$,  $\text{ellipticity} < 0.3$, and separations of at least 35 pixels (corresponding to 9.2 and 11 arcsec in DECAM and VIRCAM, respectively) were processed by \textit{PSFex}. The final photometry used the precomputed PSF model to measure the position, instrumental magnitudes, and fluxes of sources exceeding 1 standard deviation across more than 3 pixels. 

Individual zero-points were calculated as follows: first, point-like sources were selected using the \texttt{SPREAD\_MODEL} parameter, following \citet{2015ApJ...805..130K}, as shown in Equation \ref{eq:point_source}, providing robustness against SNR variations.

\begin{equation}
\label{eq:point_source}
    |\verb|SPREAD_MODEL|| < 0.003 + \verb|SPREADERR_MODEL|
\end{equation}

To avoid saturation, only sources with \texttt{MAG\_PSF} fainter than 18 ($g$ and $i$ bands) and 13 ($K_s$ band) were considered. Additionally, \(\texttt{MAGERR\_PSF} < 0.01\) mag was required to exclude low SNR sources.

An on-sky crossmatch with a 1" tolerance was performed to compute the zero point. For the $g$ and $i$ bands, the SDSS DR16 catalog \citep{2020ApJS..249....3A} was used, while UKIDSS \citep{2013yCat.2319....0L} was used for the $K_s$ band.

Finally, the zero-point was calculated as the mean difference between the external reference magnitudes and the instrumental (\texttt{MAG\_PSF}) magnitudes.

Completeness tests were carried out separately for each filter. The artificial stars used in the completeness tests were defined by a 2D Gaussian with FWHM corresponding to the specific seeing of each stacked image. The completeness value for a given magnitude was calculated by adding approximately 10000 artificial stars with magnitudes ranging between 22 and 29, and then running the same photometric pipeline as used for the data (e.g. detection, aperture photometry, PSF fitting, and final
PSF photometry), and averaging the results over 10 iterations of this procedure. We report the average magnitude at which the percentage of recovered input stars reaches $90\%$ (commonly referred to as the 90\% completeness limit). This corresponds to 24.8 mag in the $i$-band, 25.8 mag in the $g$-band, and 24.37 mag in the $K_S$ band, all in the AB system.

\begin{figure}
    \centering
    \includegraphics[width = \linewidth]{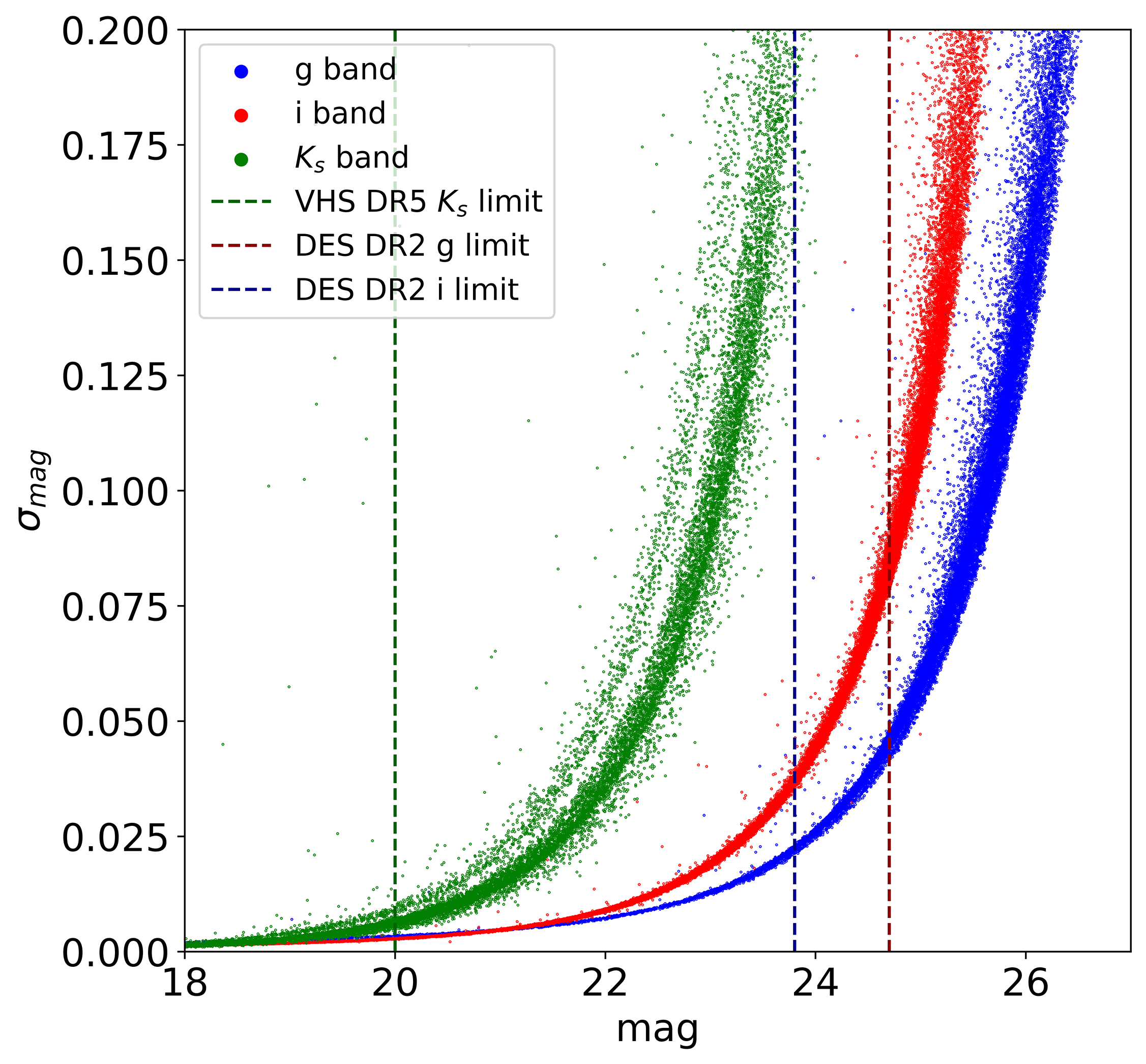}
    \caption{Photometric uncertainties of point sources in our VIRCAM/DECam stacked images as a function of magnitude. $K_s$ band uncertainties are shown in green, \textit{g} and \textit{i}-band uncertainties are presented in blue and red. The 90\% limiting magnitude for current surveys that present coverage of the region are shown as vertical dashed lines in green for the $K_s$ band VHS survey; blue and red for the \textit{g} and \textit{i} band limits in DES DR2.  }
    \label{fig:limit_comparison}
\end{figure}

Figure \ref{fig:limit_comparison} presents the photometric uncertainties for point sources detected at 1-sigma in our stacked images across all three bands (\textit{g}, \textit{i}, and $K_s$). The depth of our data is evident when comparing our photometric uncertainties with those of surveys covering the region. The 90\% completeness limits of these surveys are shown in Figure \ref{fig:limit_comparison} as dashed lines, in green for the VISTA hemisphere survey \citep[VHS;][]{2013Msngr.154...35M} $K_s$ band; in blue and red for the \textit{g}, and \textit{i} band photometry of the second data release of the Dark Energy Survey \citep[DES,][]{2021ApJS..255...20A}. The uncertainties in the $K_s$ band display two distinct sequences, resulting from a change in the noise floor of one of the stacked images, caused by the presence of a heavily saturated object.

\subsection{Extinction}

Considering the low Galactic latitude of the observed fields ( $l < -45\deg$), the extinction associated is quite low. However, there is significant differential reddening within and between each observation tile, attributable to the vast area they cover.

The extinction correction was implemented using the re-calibrated map by \citet{2011ApJ...737..103S}, originally developed by \citet{1998ApJ...500..525S} with coefficients for the \textit{g}, \textit{i} and $K_s$ bands obtained from Table 6 of \citet{2011ApJ...737..103S}.
\subsection{Star/Galaxy separation}
\label{subsec:stargal}

\begin{figure}
    \centering
    \includegraphics[width = \linewidth]{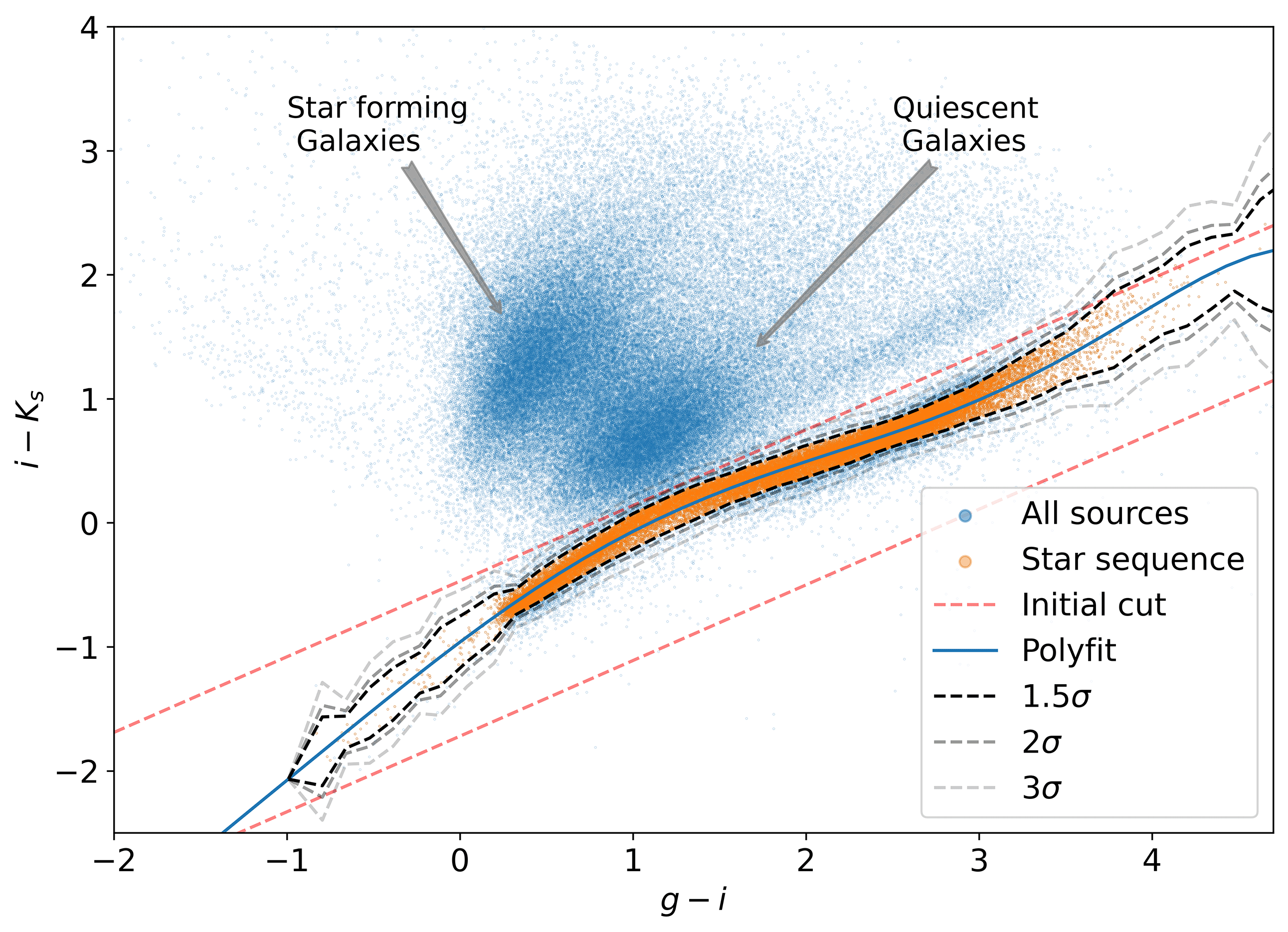}
    \caption{$g - i$ vs $i-K_s$ diagram, to which we will refer to as the $giK_s$ diagram. The stellar locus corresponds to the lower sequence stretching from ($g - i$) $\sim$ 0.0 to ($g - i$) $\sim$ 4~mag. Star selection is shown in orange and corresponds to sources found within $1.5~\sigma$ from the fitted polynomial (blue solid curve). Sources found in the group centered about $(g-i, i-K_s) = (1, 0.7)$~mag corresponds to quiescent galaxies, while the cloud centered at (0.5, 1.3) mag corresponds to star-forming galaxies.}
    \label{fig:stars_sigma}
\end{figure}

The star-galaxy separation harvests the distinct behavior of stars in the \( \textit{giKs} \) color-color diagram (Figure \ref{fig:stars_sigma}), where stellar and galaxy populations are distinguished \citep{2018ApJ...860....4O}. Star-forming galaxies appear with ($i-K_s$) $\sim$ 1.0 mag and bluer ($g-i$) colors, while passive galaxies show slightly bluer ($i-K_s$) values between 0.0 and 1.0 mag. Stars form a thin locus with bluer ($i-K_s$) colors and a wider range of ($g-i$) colors. This separation resembles the one found by \citet{2014ApJS..210....4M} using the ($u-i$), ($i-K_s$) color-color space, owing to the broad spectral coverage of the filters.

Star selection includes only point source-like detections based on morphology (satisfying Equation \ref{eq:point_source}). The selection process involves two steps: first, a broad isolation of the stellar locus is made to minimize galaxy contamination while ensuring high completeness, discarding outliers at redder ($i-K_s$) colors. The limits, shown as red dashed lines in Figure \ref{fig:stars_sigma}, are given by the polynomial: $i-K_s = 0.61 (g - i- 2) + 0.8 $ and $i-K_s = 0.61 (g - i- 2) - 0.5$. Afterward, a 6-degree polynomial is fitted to the stellar locus ( blue line in Figure \ref{fig:stars_sigma}), see \ref{eq:poly_stars} for the coefficient breakdown. Data is divided into 50 bins along the $g-i$ axis, and sources within \( 1.5~\sigma \) of the polynomial are selected as stars, and shown as orange points in Figure \ref{fig:stars_sigma}.

\begin{figure*}
    \centering
    \includegraphics[width =\linewidth]{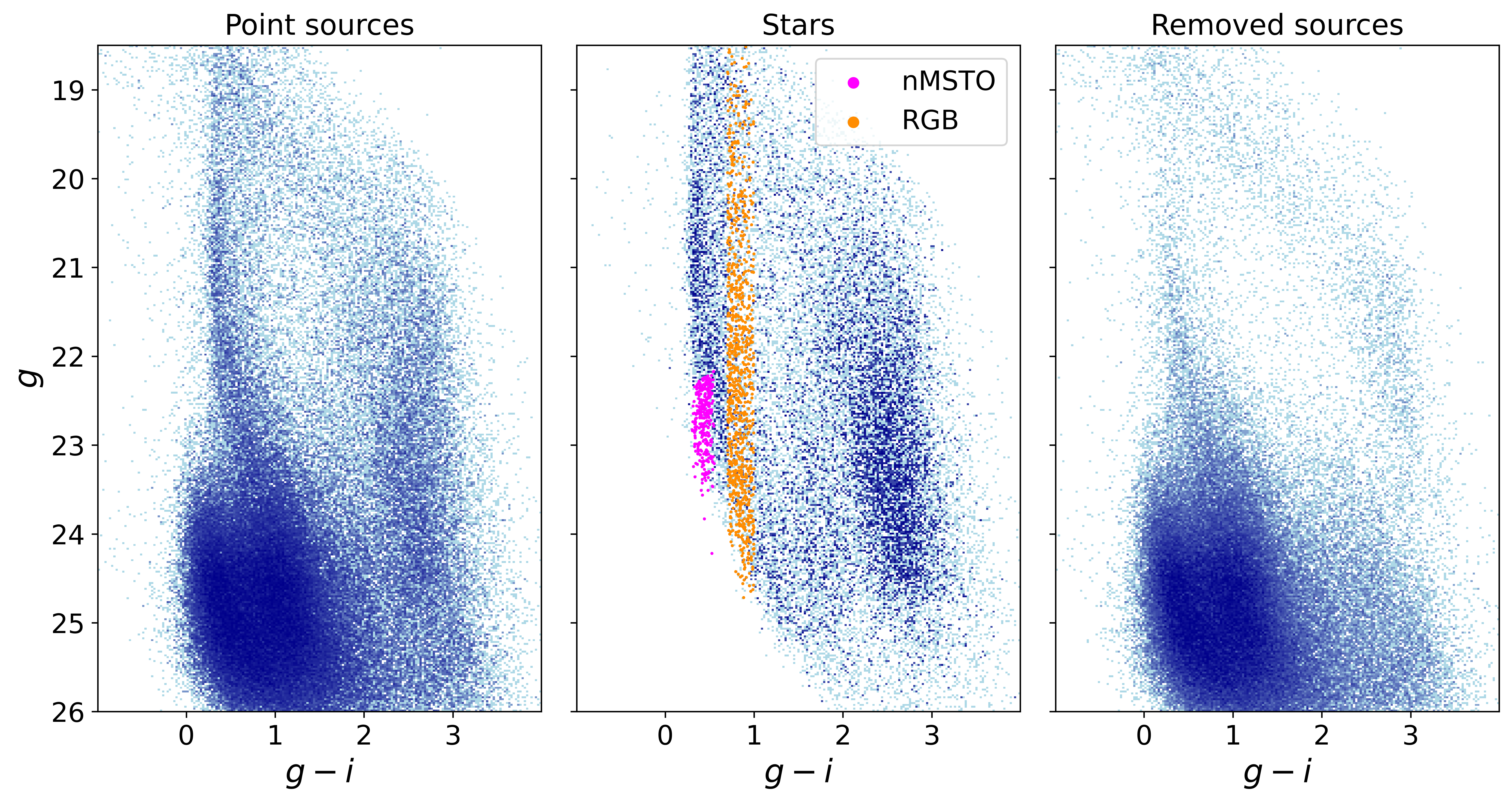}
    %\caption{Color-magnitude diagram of point sources on the left, selected stellar sources in the central plot, and removed sources are shown in the right plane. The effect of the star- galaxy separation procedure becomes evident as the removed sources corresponding mostly to non-resolved point sources like galaxies or single stellar populations occupy a large portion of the parameter space. }
    \caption{The left panel presents the color-magnitude diagram of point sources in blue. The central plot highlights the selected stellar sources. The nMSTO and RGB samples are shown in magenta and orange respectively. The right panel shows the sources that were removed through the star-galaxy separation procedure applied to point sources. The efficacy of the star-galaxy separation procedure is evident in this arrangement. The removed sources, predominantly comprising non-resolved sources such as galaxies or single stellar populations, occupy a significant portion of the parameter space and dominate the counts at $g \geq 22$, thus overlapping the nMSTO and RGB selection at large distances.}
    \label{fig:cmd_stargalax}
\end{figure*}

The results of this procedure are shown in the Hess diagram (i.e., the density of sources in the color-magnitude space) in Figure \ref{fig:cmd_stargalax}: the left panel presents point sources from all tiles amounting to 170\,635 objects, where an increased number of sources have $g-$band magnitudes fainter than 23~mag, stressing the need for an effective star/galaxy separation. After removing galaxies as described above, in the middle panel we show that there are 31\,872 stars from all tiles, where the effect of non-resolved galaxies becomes evident as they occupy the same color and magnitude space as stars in the CMD; the right panel shows the sources removed as a result of the star/galaxy separation procedure, which corresponds to 138\,763 objects (i.e., $\sim 81.3\%$ of the total point sources). 

\section{Halo tracers} \label{sec:tracers}

\subsection{Red Giant Branch Stars}
\begin{figure}
    \centering
    \includegraphics[width = \linewidth]{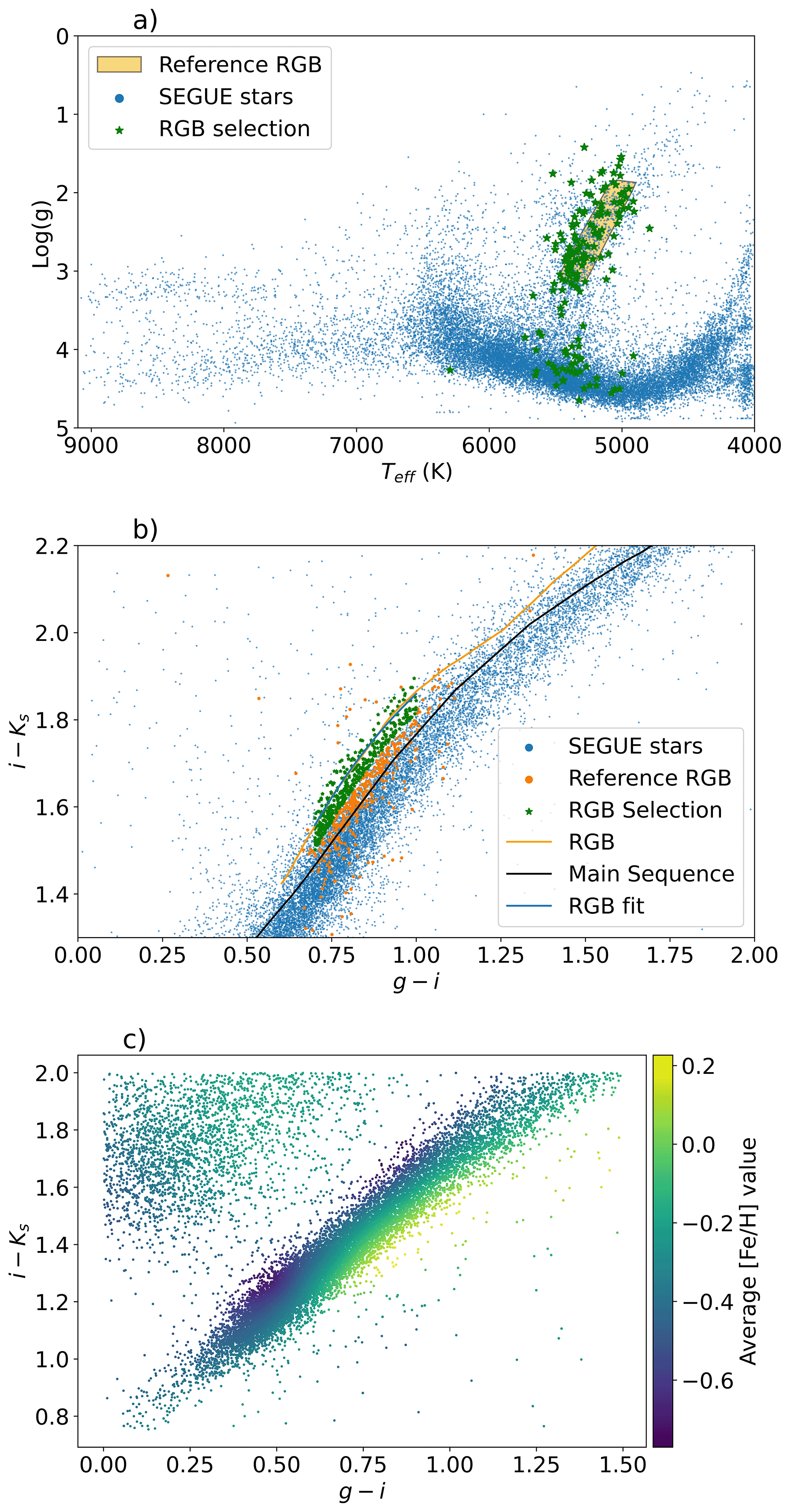}
    \caption{a)  Kiel's diagram/spectroscopic HR diagram shows the logarithm of surface gravity ($\log(g)$) as a function of the effective temperature ($T_{\text{eff}}$) for SEGUE-2 stars using SDSS and UKIDSS photometry. The adopted selection of RGB stars is shown as an orange polygon (see the main text). Stars meeting the RGB selection criteria in the giKs diagram are marked as green stars, highlighting the main sequence contamination evident from 23.1\% of stars with $\log(g) > 4$ within the RGB selection. b) The $giK_s$ color-color diagram displays all stars in blue, giants selected on the Kiel diagram (orange polygon in the left panel), and the RGB selection in green. The RGB and main sequence for a 10 Gyr, [M/H] = -1.5 [M/H] isochrone are depicted in orange and black, respectively. c) $giK_s$ color-color diagram of LAMOST DR8 stars having DES and UKIDSS photometry, colored by metalicity showing the metallicity gradient present in the stellar locus, allowing for the separation of low metalicity ($[M/H] <$ -0.6 dex) RGB stars from the main sequence. A robust locally weighted regression has been applied to display the broad tendency and reduce noise. }
    \label{fig:combined_vertical}
\end{figure}

To select red giant branch (RGB) stars using \textit{g}, \textit{i}, and $K_s$ band photometry, we first identified evolutionary sequences in the $giK_s$ color-color diagram. Data from the Sloan Extension for Galactic Understanding and Exploration 2 \citep[SEGUE,][]{segue}, which provides low-resolution spectra (R$\sim 1800$) for 118,958 stars, was used to guide the selection. SEGUE prioritizes distant halo tracers such as BHBs, K giants, and M giants and provides derived fundamental parameters, enabling the clear selection of RGB stars in the Kiel diagram (Figure \ref{fig:combined_vertical}a).

We defined a reference sample to which we will refer to as the SEGUE RGB sample. These stars were selected according to their position in the Kiel's diagram to follow the center of the RGB locus, this was implemented by selecting sources within the polygon defined by the points: $(T_{ \rm eff}, log(g))$ = (4923, 1.98), (4894, 1.87), (5039, 1.84), (5088, 1.88), (5216, 2.25), (5294, 2.48), (5317, 2.55), (5399, 2.81), (5418, 2.87), (5460, 3.02), (5467, 3.04), (5467, 3.06), (5451, 3.11), (5278, 3.14), highlighted as the orange selection in Figure \ref{fig:combined_vertical}a. %This selection was designed to minimize contamination, focusing on the center of the RGB locus.

We crossmatched SEGUE stars with \textit{g}, \textit{i} photometry from SDSS and $K_s$ photometry from UKIDSS DR9 \citep{2013yCat.2319....0L}, enabling us to map them in the $giK_s$ diagram (Figure \ref{fig:combined_vertical}b). The SEGUE RGB sample forms a sequence in $giK_s$ space, shown as orange dots in Figure \ref{fig:combined_vertical}b.

By coloring SEGUE stars by metallicity in the $giK_s$ diagram, we observed that low-metallicity stars ($[M/H] < -1.3$) shift towards higher $i-K_s$ colors. To confirm this feature is not unique to SEGUE, we crossmatched LAMOST DR8 stars with SDSS and UKIDSS photometry, then plotted them in the $giK_s$ diagram, colored by metallicity. Applying a locally-weighted regression \citep{2013MNRAS.432.1862C, doi:10.1080/01621459.1988.10478639} highlighted a clear metallicity gradient, trending diagonally towards redder $(i - K_s)$ and bluer $(g - i)$ colors. This allows the separation of low-metallicity RGB stars from the main sequence using only the $g$, $i$, and $K_s$ bands, suggesting a correlation between $i-K_s$ color and stellar metallicity.

Guided by the SEGUE RGB reference sample, we selected RGB stars from our VIRCAM/DECAM data using the criteria: $0.7 \leq g - i \leq 1$ mag (matching SEGUE RGB stars in Figure \ref{fig:combined_vertical}.b) and a perpendicular distance $<0.035$ mag from a PARSEC \citep{2012MNRAS.427..127B} isochrone (10 Gyr, [M/H] = -1.5 dex). This ensures separation from the main sequence and captures stars with halo metallicities ($-1.3$ to $-2.3$ dex).

To remove foreground sources, we crossmatched with Gaia DR3 \citep{2023A&A...674A...1G} and applied a parallax cut ($\pi < 0.1$ mas, distances $>10$ kpc), filtering out 327 bright foreground objects from 1261 initial sources.

We validated our RGB selection by applying it to SEGUE-2 stars in the $giK_s$ diagram and confirmed their RGB status in the Kiel diagram (green dots in Figure \ref{fig:combined_vertical}.b and green stars in Figure \ref{fig:combined_vertical}.a). Excluding stars with $log(g) > 4$ as main sequence contaminants \citep{1996BABel.154...13A}, we achieved a purity of 77.9%, with 23.1% contamination.

Applying this selection to our photometric catalog yielded 934 giants across 4 fields, or $\sim 142 \text{ stars/deg}^2$.

\subsection{Near main sequence turn off stars}

The second tracer available in our photometry is Near Main Sequence Turn-Off (nMSTO) stars, which are significantly more abundant than other typical halo tracers, such as RR Lyrae, BHB stars, and K giants. nMSTO stars are the faintest halo tracers reached by our photometric catalogs. Considering the completeness limit of our data, such stars can be identified up to a distance of 100 kpc. This tracer has been widely used to probe the outer halo \citep{ 2001ApJ...553..184C, 2008ApJ...673..864J, 2008ApJ...680..295B, 2011ApJ...731....4S} as their larger abundance allows to place competitive limits on the global density profile and shape of the Galactic halo with a relatively small number of narrow fields of view \citep{2015A&A...579A..38P}.

To select nMSTO stars, we used the selection criteria performed in \citet{2015A&A...579A..38P} as a reference to device a selection method using the \textit{g}, \textit{i}, and $K_s$ bands. Their selection is based on the \textit{u, g, r,} and \textit{i} photometry, by applying two empirical photometric variables. The first is the metallicity [Fe/H], which is determined using \cite{2010ApJ...716....1B} photometric metallicity relation based on SDSS spectroscopic metallicity. The second empirical variable is the absolute magnitude $M_r$, derived from \citet{2008ApJ...684..287I} photometric parallax relation, which was calibrated from globular clusters with known distances in SDSS data. The resulting color limits are shown in equations 9 to 13 in \cite{2015A&A...579A..38P}.

\begin{figure}
    \centering
    \includegraphics[width = 0.99\linewidth]{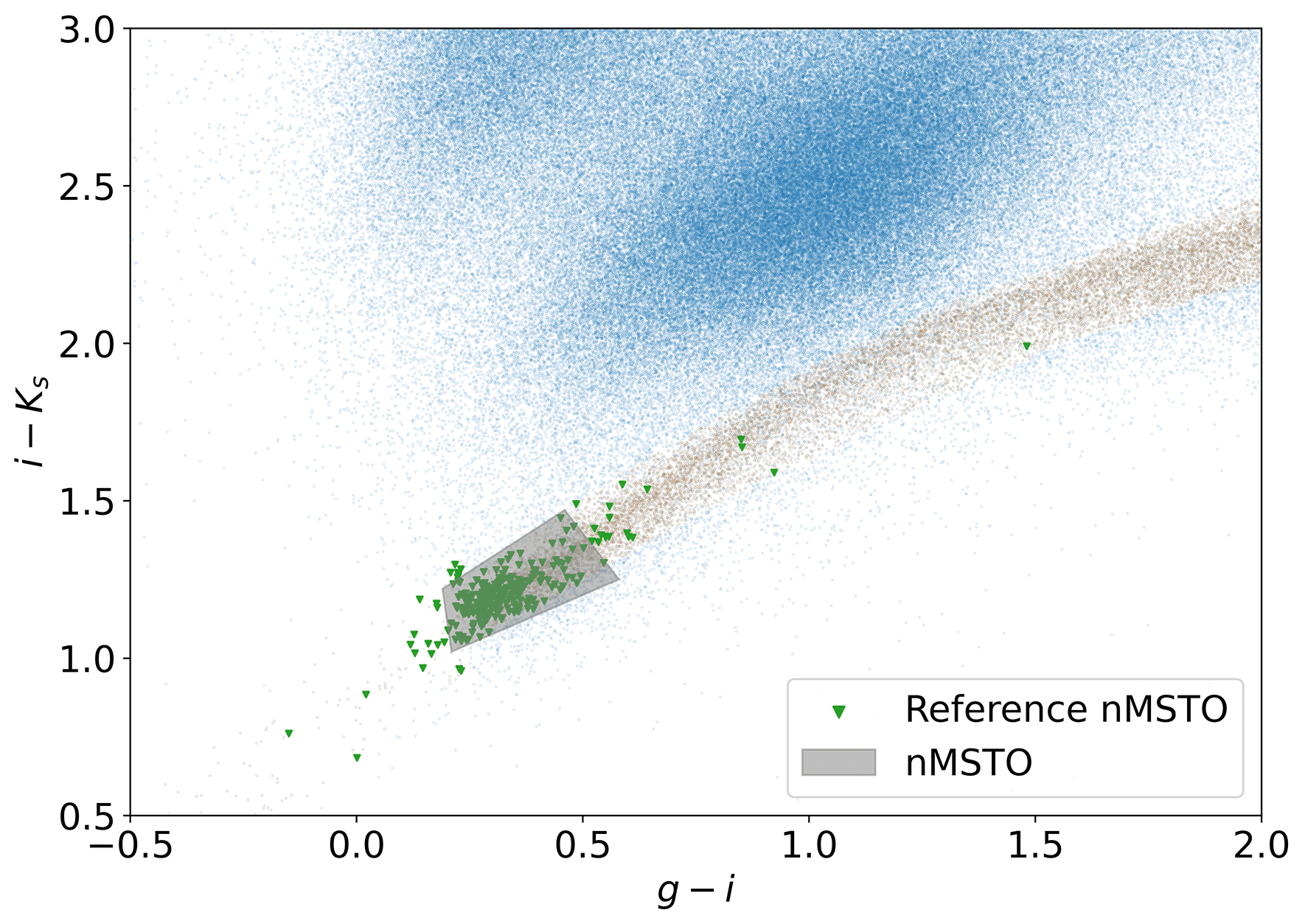}
    \caption{$giK_s$ diagram showing point sources in blue, while stars shown in orange. Green triangles correspond to the reference selection of nMSTO stars obtained by applying \cite{2015A&A...579A..38P} criteria. Our polygon for defining nMSTO stars is shown as a grey box.}
    \label{fig:nMSTO}
\end{figure}

By applying  \citet{2015A&A...579A..38P} criteria on the photometry from DES we obtained a reference sample of nMSTO stars. The use of DES photometry is essential, as it provides photometry across all the bands used by \cite{2015A&A...579A..38P}. This reference sample is then matched with UKIDS photometry by a crossmatch with a tolerance of 1 arcsecond. This provides a sample of nMSTO stars with $giK_s$ photometry, which we show in Figure \ref{fig:nMSTO} as green triangles. 

The selection box for nMSTO stars in our photometry is given by the following box:
($g-i$, $i-K_s$) = (0.21, 1.02), (0.19, 1.22), (0.46, 1.47), (0.58, 1.25). This box was selected to include the majority of the reference sample while excluding outliers, this implies avoiding BHBs or Blue Stragglers, which are expected to be found beyond $g - i < 0.2$. Additionally, we excluded stars with $g - i > 0.6$ since they are no longer near the main sequence turn-off. Contamination is expected to be low within this box since in the $giK_s$ color-color diagram old (10 Gyr or more) isochrones have their turn-offs at the tip of the stellar locus, with the subgiant branch and RGB being `folded' towards redder regions of this color space.  Utilizing this method we obtained a sample of 2\,554 nMSTO stars among the 4 fields, which corresponds to $\sim 388 \text{ stars/deg}^2$, doubling the abundance of our RGB sample.

\subsection{Distances} \label{sec:dist}

Photometric distances for both tracers were estimated by employing PARSEC \citep{2012MNRAS.427..127B} isochrones. The selection of isochrones must cover both the metallicity distribution function and the age ranges that are expected to be found in the MW halo.

Particularly for the MW halo, \citet{2013ApJ...763...65A} found the metallicity distribution function to peak at  $[M/H] \simeq -1.7$ and $-2.3$. Alternatively, \citet{2019ApJ...887..237C} found that the mean halo metallicity to be $[Fe/H] \simeq -1.2$, displaying no gradient between $6-100$ kpc.

Since our RGB selection imposes an upper limit in the metallicity of $[M/H] \simeq -1.3$, we will apply a flat halo prior that includes isochrones with a metallicity range of $[M/H] \simeq -1.3$ to $-2$. The ages pose a secondary role, as metallicity has a larger impact on the inferred distances \citep{2021Natur.592..534C}. For completeness, we chose to include isochrones with ages from $10-12$~Gyr, which takes into account most of the MW halo age distribution \citep{2022Natur.603..599X, 2011A&A...533A..59J}.

We utilize the (g - i) color of each tracer star to interpolate its absolute magnitude from the isochrones in the previously described grid. Using the observed apparent g magnitude, we then derive the distance modulus between each tracer and each isochrone in the grid described above. We calculate the distance for each tracer as the average distance modulus across the entire isochrone grid, utilizing a flat prior within the metallicity and age range described above. The distance error corresponds to the $1\sigma$ standard deviation across the grid. The typical distance error for nMSTO and RGB sources is ~ 10\% and ~ 25\% respectively. We emphasize that precise distances are not required for our analysis. 

Utilizing the previously computed distances, we selected a total of 102 giants and 309 nMSTO stars between 60 - 100 kpc in Galactocentric distance, that will be used in subsequent analysis, for a full breakdown in line of sight, see table \ref{tab:star_counts}.  Of the RGB sample 5, 17, 32 and 48 are found in Tiles 1, 2, 3 and 4 respectively, for the nMSTO sample 44, 54, 94 and 117 are found in Tiles 1, 2, 3, and 4 respectively. 

\begin{deluxetable}{ccc}
\tablecaption{Number of stars found between 60-100 kpc in each tile for each tracer.\label{tab:star_counts}}
\tablehead{
\colhead{Field} & \colhead{nMSTO} & \colhead{RGB}
}
\startdata
Tile 1 & 44 & 5 \\
Tile 2 & 54 & 17 \\
Tile 3 & 94 & 32 \\
Tile 4 & 117 & 48 \\
\enddata
\end{deluxetable}

\section{Simulations details} \label{sec:sim}

To serve as a point of comparison, and to plan the coordinates of the observed fields in this work (see Figures \ref{fig:observations_plot} and \ref{fig:lmc25_ra}) we utilized the high-resolution N-body simulations of the interaction between the Milky Way (MW) and the Large Magellanic Cloud (LMC) presented by \citet{2019ApJ...884...51G}. These simulations model the MW's dark matter halo with a Hernquist profile and a virial mass of \(1.2 \times 10^{12} M_\odot\), accompanied by a stellar disk and bulge with masses of \(5.78 \times 10^{10} M_\odot\) and \(0.9 \times 10^{10} M_\odot\), respectively. The particle mass for dark matter in these simulations is \(4 \times 10^4 M_\odot\), offering high-resolution results that capture the interaction between the MW and the LMC.

For the LMC, we chose the model with a virial mass at infall of $2.5 \times 10^{11} M_{\odot} $, as it generates the largest wake signal and better aligns with the observations in our study. The LMC's dark matter halo is modeled with a Hernquist profile. The orbital parameters of the LMC are derived iteratively to ensure that its present-day position and velocity are within $2\sigma$ of observed values. The MW-LMC interaction is modeled over the past 2 Gyr, including dynamical friction effects from the LMC's passage.

Two MW halo models are considered by \citet{2019ApJ...884...51G}: one with isotropic halo kinematics and the other with a radially biased kinematic profile. These models allow for the exploration of the effects of different halo kinematics on the resulting dynamical friction wake and perturbations in both the stellar and dark matter components of the halo. For a full description of the simulation setup and detailed parameters, we refer the reader to \citet{2019ApJ...884...51G}.

\section{Stellar density profiles} \label{sec:results}

\subsection{Density variations across the wake} \label{sec:ra_dens}

All of our four fields are aligned at DEC = 0$^\circ$, hence we can present a projected density profile by comparing the stellar counts for each tracer (nMSTO and RGB stars) with respect to their Right Ascension (RA). Based on \citet{2019ApJ...884...51G} simulations, the range of RA spanned by the 4 observed fields will cover the full density range that is expected to be found for the wake stellar overdensity. 

To define a zero point for the density profiles, we first compute the expected mass of the Milky Way halo between 60 and 100 kpc, where the density contrast and therefore the signature of the wake is expected to be the highest \citep{2019ApJ...884...51G}. This mass is obtained by integrating the best-fit radial density profile for the MW halo according to \citet{2015ApJ...809..144X}, which corresponds to an Einasto profile \citep{1965TrAlm...5...87E} with a concentration index of $n = 3.1 \pm 0.5$, with an effective radius $r_{\rm eff} = 15 \pm 2$ kpc and a flattening of $q = 0.7 \pm 0.02$ defined as follows:

\begin{equation}\label{eq:einasto}
    \rho_{*}(r_q) \equiv \rho_0 \exp[{-d_n[\left( r_q/r_{eff}\right)^{(1/n)} - 1]}]
\end{equation}

\noindent where $d_n \approx 3n - 1/3 + 0.0079/n$ ; $r_q = \sqrt{R^2 + (z/q(r))^2}$ is the basic Galactocentric radial coordinate given the flattening $q(r)$,  the Galactocentric radius $R$ and the Galactic coordinate $z$ ; $\rho_0$ corresponds to the local normalization parameter.

Integrating Equation~\ref{eq:einasto} using \citet{2019MNRAS.490.3426D} local normalization of $\rho_0 = 6.9 \times 10^{-5}M_{\odot}pc^{-3}$, we get an outer halo stellar mass between $60 - 100$~kpc of $ \approx (1.8 \pm 0.4) \times 10^{8} M_{\odot}$. This mass estimate includes stars of all evolutionary states. Therefore, a mass-to-number ratio for nMSTO and RGB stars is needed to compute an expected average number of sources for each of our tracers, which will be used later to obtain an average background stellar halo density with respect to which observed density contrasts will be computed and discussed.

\begin{figure*}
    \centering
    \includegraphics[width = 0.6\linewidth]{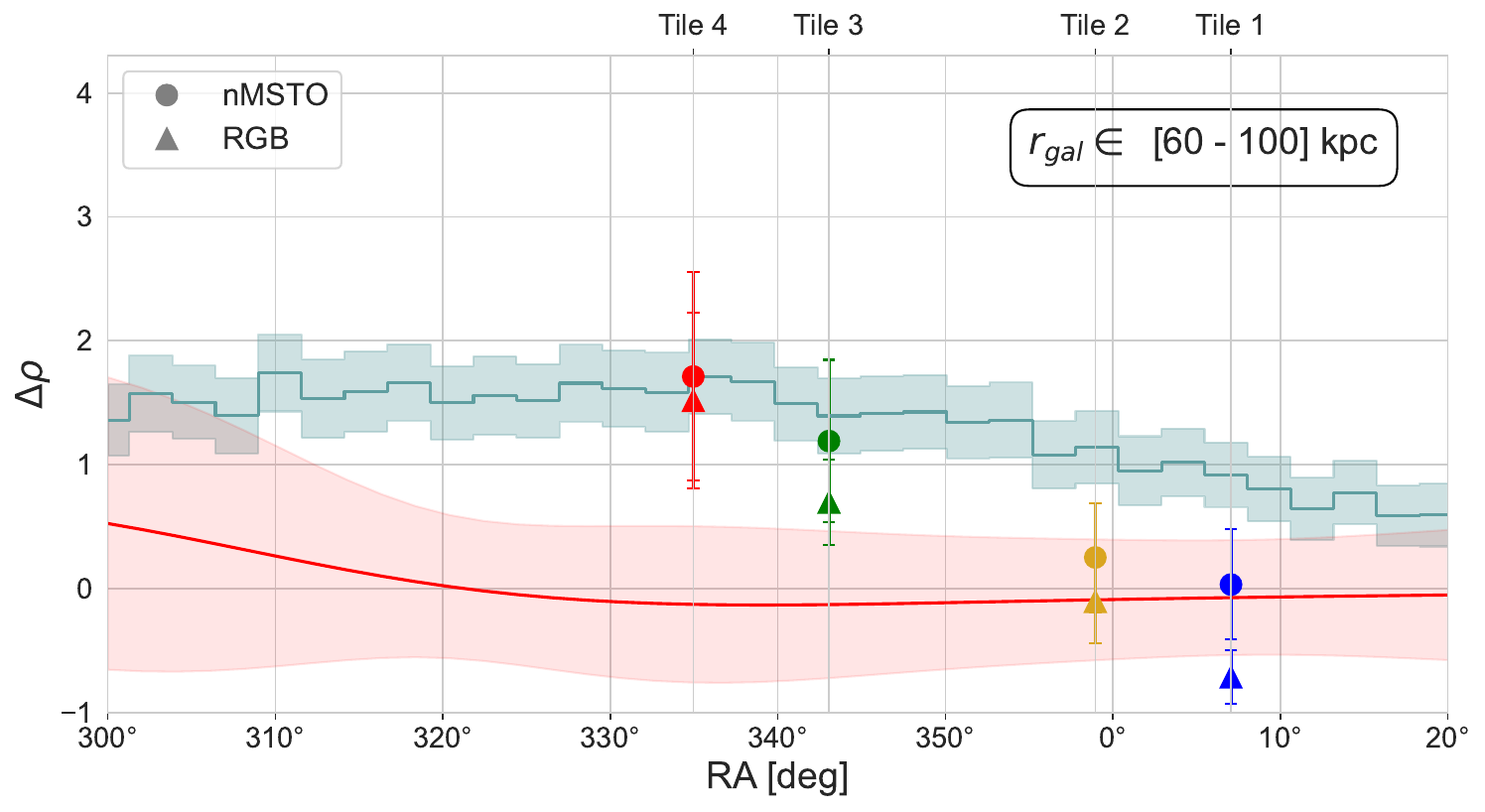}
    
    \caption{Derived density profile $\Delta \rho$, integrated along the line-of-sight between 60-100 kpc from the Galactic center and projected as a function of RA. The number density of nMSTO stars is indicated as points, and the density of RGB stars is shown as upward pointing triangles, for both the color is dependant on the corresponding Tile, with red, green, yellow, and blue for Tiles 1, 2, 3 and 4. Both tracers are normalized according to the best-fit Einasto profile by \citet{2015ApJ...809..144X}, as described in the text. The light blue histogram represents the number density profile of dark matter for the same segment of the halo between 60-100 kpc, from the \citet{2019ApJ...884...51G} model including the formation of a wake owing to the passage of the LMC with a virial mass at infall of $2.5\times 10^{11}$ M$_{\odot}$, normalized by the average density ($\overline{\rho}$) in the entire sky of the simulation. Error bars, shaded in light blue, reflect 1000 bootstraps of the slice, encompassing the same DEC as the observed fields. A kernel density estimate of the average relative number density profile in RA for three random DEC coordinates in all \citet{2005ApJ...635..931B} halos is shown in red, with the surrounding lighter red area corresponding to the $1 \sigma$ dispersion. A clear density enhancement is observed for both nMSTO and RGB stars, peaking at Tile 4, in line with the simulation.}

    \label{fig:lmc25_ra}
\end{figure*}

Using PARSEC 1.2s tracks \citep{2012MNRAS.427..127B}, we simulated a 10 Gyr-old stellar population with a metallicity of -1.5 dex, corresponding to the peak of the halo metallicity distribution function and age distribution. This simulation drew from a Kroupa initial mass function, corrected for non-resolved binaries \citep{2001ASPC..228..187K, 2013pss5.book..115K}, and a total mass of $10^{5}$ M$_{\odot}$ given the limits of the code. We can then obtain a mass-to-number ratio by applying our color selection criteria presented in Section \ref{sec:tracers}, to the simulated photometry output in SDSS $g$, $i$ and 2MASS $K_s$ bands and dividing the resulting number of stars by the known total mass of the simulated population. This procedure yielded mass-to-number ratio estimates of 149.7 M$_{\odot}/{\rm star}$ for nMSTO stars and \(343.4 M_{\odot}/{\rm star}\) for RGB stars. Given these ratios and the previously determined stellar mass of the halo between $60-100$~kpc, we get an estimated $19 \pm 5$ giants and $43 \pm 11$ nMSTO stars for each field, which will represent the average stellar halo background density (i.e., $\overline{\rho}$, see Equation \ref{eq:deltarho}) in Figure \ref{fig:lmc25_ra}. This accounts for roughly $2/3$ of our observed sample between 60-100 kpc, which includes 102 giants and 309 nMSTO stars.

The previously described process for computing $\overline{\rho}$ allows for a rough estimation of the real all-sky density. The corresponding caveats and assumptions involved in this process are discussed in section \ref{sec:assumptions}. 

Figure \ref{fig:lmc25_ra} illustrates the density profile in RA at DEC$=0^\circ$, integrated between the line of sight and considering only stars found between 60 and 100 kpc from the Galactic center. The changes in the local density ($\Delta \rho$) are measured for each tracer as shown in equation \ref{eq:deltarho}, where the measured number of stars between 60 and 100 kpc ($\rho$) is divided by the previously computed average number density ($\overline{\rho}$) of the entire MW halo, according to the best-fit Einasto profile by \citet{2015ApJ...809..144X}. Errors represent the $1\sigma$ limit obtained after 1000 iterations, during which distances for individual stars are resampled under the assumption of Gaussian distributions centered on the mean distance, with standard deviations defined by the respective distance uncertainties, and the density is recomputed. The dark matter particle number density of the \citet{2019ApJ...884...51G} simulation, including the formation of a DM wake owing to the first infall orbit of a $2.5\times 10^{11}$ M$_{\odot}$ LMC, is presented in light blue and normalized to the $\overline{\rho}$ of the entire halo of the simulation. Error bars, shown in light blue, reflect 1000 bootstraps of the slice using 10\% of the particles in each iteration, encompassing the same DEC as the observed fields.

\begin{equation}\label{eq:deltarho}
    \Delta \rho =  \rho/\overline{\rho} - 1\text{.}
\end{equation}

Figure \ref{fig:lmc25_ra} presents a clear overdensity, consistent between both tracers, showing the maximum measured density in Tile 4 at RA $\sim 335$ deg with an average overdensity of $\Delta \rho = 1.6 \pm 0.6$. The lowest density is measured for Tile 1 at RA $\sim 7$ deg for both nMSTO and RGB stars. Although both tracers exhibit a similar trend of decreasing density from Tile 4 to Tile 1, the densities measured by each tracer become inconsistent at the $1\sigma$ level in Tile 1, likely due to the low number of stars present in the RGB sample (5 stars) in Tile 1.

We perform a Z-test \citep{DictionaryOfStatistics2008} to probe the significance of these variations in density as a function of the RA coordinate. The null hypothesis is that of a hierarchically formed spherical halo, without the influence of the LMC. This hypothesis is grounded in the concept that Milky Way-like galaxies evolve through the hierarchical growth process, involving mergers and accretion of smaller systems. These systems, once influenced by the MW's tidal gravitational field, become progressively more mixed. This process eventually leads to the formation of a spheroidal component of the halo, a result of their collective assembly, with a significant portion of it constituted by inhomogeneous substructures such as streams, plumes, and dwarf satellite galaxies \citep{1978ApJ...225..357S, 2005ApJ...635..931B, 2006MNRAS.365..747A, 2012AN....333..460D, 2015AJ....150..128R, 2019MNRAS.485.2589M}.

For this purpose, we look at the simulated halos of \citet{2005ApJ...635..931B}, which are entirely hierarchically constituted. Since in their simulations, the halo beyond $\sim 60$ kpc is spherical, we considered 3 random stripes with a major axis equal in length to the angular range shown in Figure \ref{fig:lmc25_ra} and a minor axis equal to the angular size of a VIRCAM field of view \citep[1.64 deg$^2$][]{2006Msngr.126...41E}, and selected stellar particles between 60 and 100 kpc from the Galactic center. The mean kernel density estimate for $\Delta \rho$ is almost constant around zero and the mean standard deviation found across all 11 \citet{2005ApJ...635..931B} halos is on average $\overline{\sigma_{bj}} = 0.59$. Therefore, the null hypothesis for the Z-test is that the observed relative densities follow a flat distribution with a mean of zero and a standard deviation of $0.59$. The Z-score for the highest density field (Tile 4) is 2.55 and 2.88 for RGB and nMSTO stars respectively, which corresponds to an average p-value of $0.003$. Thus passing the 95\% significance test. Note that this value is directly dependent on the chosen profile and local normalization used to define $\overline{\rho}$. This caveat is discussed in section \ref{sec:assumptions} %See table\ref{tab:zscore_combined} for a breakdown of the z-test on each tile and tracer. % Tabla comentada para ahorrar dinero

\subsection{Radial Density Profiles}\label{sec:radial}

An analysis of the radial density profile facilitates the detection of substructures along the line of sight \citep{2015A&A...579A..38P, 2015ApJ...809..144X}. Additionally, \citet{2019ApJ...884...51G} predicts a measurable footprint of the wake in the radial density profile.

To construct the radial density profile for the more abundant nMSTO stars, they are split into 8 bins by their Galactocentric radius between 20 - 100 kpc, we then calculate the number density per bin and line of sight ($\rho_{l, b, r} $) as follows:

 \begin{figure*}[ht]
     \centering
     \includegraphics[width = \linewidth]{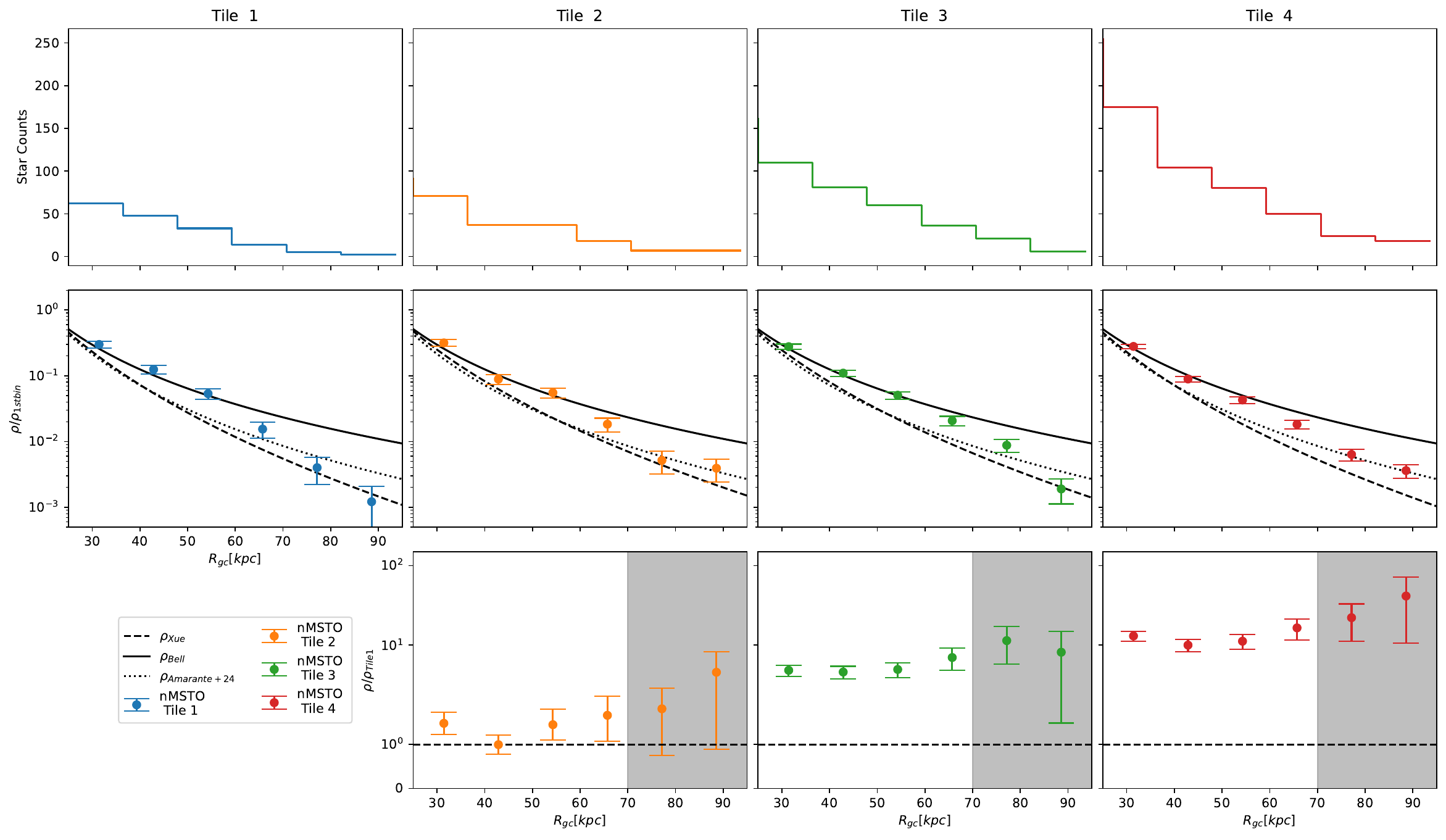}
     \caption{First-row shows the star counts as a function of galactocentric radius for nMSTO stars. The second row shows the stellar density profiles for the nMSTO stars for the 4 observed fields (ie., the first row divided by volume), normalized according to the number density of the highest density bin at 25 kpc (just after the expected position of the break in the halo), thus bypassing the computation of an error-prone normalization parameter for each radial density profile and allowing comparison in the slope. \citet{2015ApJ...809..144X}, \citet{2008ApJ...680..295B} and \citet{amarante24} radial density profiles are shown as dashed, continuous and dotted black lines respectively}. nMSTO radial density is shown in blue, orange, green, and red corresponding to Tiles 1,2,3, and 4, this holds for the entire plot. The third row shows the radial density profiles relative to Tile 1. Error bars correspond to Poisson errors. We find our data closely resembles \citet{2008ApJ...680..295B} radial density profile up to $\sim$ 70 kpc where \citet{2015ApJ...809..144X} profile shows a better agreement. There is an increase in density with respect to Tile 1 that increases further at 70 kpc.% 2, 3, and 4, with Tile 1 showing an excess at $\sim$ 85 kpc indicative of substructure contamination. }
     \label{fig:radial_b1_25}
\end{figure*}

\begin{equation}
    \rho_{l, b, r} = \frac{N_{l,b,r}}{V}
\end{equation}
\noindent Where $V$ corresponds to the volume contained within each galactocentric bin ($r_1, r_1+ 7.27$ kpc) computed by integrating the solid angle of the observed field with the corresponding heliocentric distance $r_h$ that results in the corresponding bin.

The results of the number density calculations are presented in Figure \ref{fig:radial_b1_25}, where the nMSTO stellar counts are shown as a function of the Galactocentric distance $R_{gc}$ for each Tile (shown in the upper row of panels of Figure \ref{fig:radial_b1_25}).  These number counts are then divided by the volume of each bin and normalized to the maximum density bin (which corresponds to the bin at 25 kpc), thus producing the radial density profile shown in the middle row of panels of Figure \ref{fig:radial_b1_25}. The normalization allows for a comparison between the shapes of published stellar halo profiles and our data, as each profile requires a scaling factor to fit the data. For reference, recall that Tile 1 (leftmost panel) corresponds to the lowest density measured along our pointings, and Tile 4 (rightmost panel) to the peak density, for tracers between $60 - 100$~kpc (Figure \ref{fig:lmc25_ra}). The triaxial broken-power law density profile from \citet{2008ApJ...680..295B} is shown as a black curve, along with the Einasto profile from \citet{2015ApJ...809..144X} as a dashed line, and the best fit profile found for the Pisces region in \citet{amarante24} as a dotted line.
 % labeled as $\rho_{\rm Bell}$ and $\rho_{\rm Xue}$, respectively. 
Figure \ref{fig:radial_b1_25} illustrates that the density profiles decrease as a function of Galacticentric radius in all tiles, consistent with \citet{2008ApJ...680..295B} triaxial double power-law model only up to ~ $60-70$ kpc.  Beyond this point, a discontinuity and change in slope is observed, where our data appears to be better explained by either \citet{2015ApJ...809..144X} Einasto profile or \citet{amarante24} fit for the Pieces region. 

\begin{figure}
    \centering
    \includegraphics[width=\linewidth]{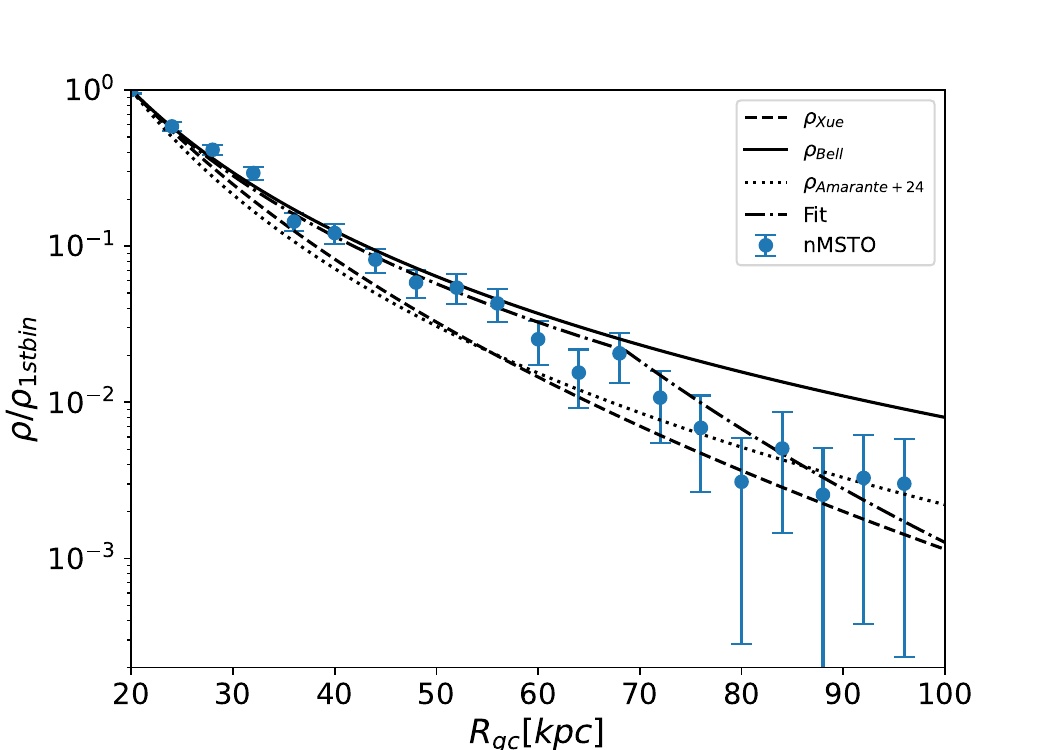}
    \caption{The combined radial density profile for all four tiles shows the nMSTO number density normalized by the density of the first bin, enabling a comparison of slopes. The radial density profiles from \citet{2015ApJ...809..144X}, \citet{2008ApJ...680..295B}, and \citet{amarante24} are represented by a dashed line, solid line, and dotted line, respectively. Our double power law fit for the nMSTO radial profile is depicted as a red dot-dashed line.  The break that was apparent in the middle row of panels of Figure \ref{fig:radial_b1_25} is now significantly more clear, with best-fit value at $67.5^{+4.8}_{-5.0}$ kpc. }
    \label{fig:fit}
\end{figure}

The apparent deviation from the Bell profile in all tiles beyond $\sim$ 60 kpc is very suggestive of a real change, as this happens to be the distance regime at which the LMC wake is expected to manifest \citep{2019ApJ...884...51G}. This change in the slope of the halo density profile after $\sim$ 60 kpc becomes very clear in Figure \ref{fig:fit}, where we stack the data of all our pointings, as discussed ahead.

From Figure \ref{fig:observations_plot}, our fiducial simulation predicts a density gradient across the 4 tiles, where Tile 1 should be the least dense relative to the other tiles, and Tile 4 should be the most dense. This was confirmed in projection in Figure \ref{fig:lmc25_ra}.  We now examine the behavior of this relative density enhancement as a function of Galactocentric distance. We achieve this by dividing (bin by bin) the observed stellar densities of Tiles 2, 3, and 4 by the corresponding ones observed for Tile 1.  This produces, as shown in the bottom row of panels of Figure \ref{fig:radial_b1_25}, observed halo density profiles relative to a field in the outskirts of the wake, for three adjacent sight lines that go, across the sky, into the expected location of the peak of the LMC wake.  The trend of increasing density from Tile 2 to 4 now appears clearly.  Moreover, in Tiles 2, 3, and 4 there are hints of a jump in the stellar density relative to Tile 1 that starts at $\sim$ 70 kpc and that increases in magnitude across the sky, peaking in Tile 4. Such an increasing density gradient across the sky (Tiles 1 to 4) and that jumps starting at $\sim$ 70 kpc is consistent with the predictions from the simulations of \citet{2019ApJ...884...51G}. Importantly, the angular scale over which our density profile displays the density jump (Tile $1 - 4$ = 32.5 degrees) is larger than the expected angular scale subtended by substructure in hierarchically constituted halo simulations (see \ref{subsec:substructure}).

Additionally, for Tiles 3 and 4 we observe an increasing density contrast at small distances as well (between 20 - 70 kpc), which maintains the same $\rho_{\rm Bell}$ slope but at higher star counts. This behavior indicates that the halo in this region of the sky may be highly anisotropic and thus non-spherical (see \citealt{triaxial2024}). Alternatively, within 70~kpc this could be due to the presence of contamination from Sagittarius, even though our fields are at higher latitudes than where the main track from the Sagittarius stream is.

Finally, to confirm or reject the reality of a new density break observed in all Tiles between $\sim 60-70$ kpc and better determine its location, we combined the data from all 4 Tiles to boost the star counts, which allows the construction of a density profile with a better spatial resolution (i.e., with bins of 2.5 kpc instead of 10 kpc).  We fit this combined radial density profile with a triaxial double power law using a Monte Carlo Markov Chain (MCMC) approach, which yields a best fit of $\alpha_{in} = 3.13^{+0.48}_{-0.61} $, $\alpha_{out} = 7.46^{+0.44}_{-0.53}$, with a break radius of $67.5^{+4.8}_{-5.0}$ kpc, and presented in Figure \ref{fig:fit} with a dot-dashed red line. We additionally fit the triaxiality ratios, for which we found $b/a = 1.0^{+0.57}_{-0.44}$ and $c/a = 0.98^{+0.49}_{-0.50}$.For a detailed breakdown of the fitting procedure see \ref{sec:mcmc}. %Since the data from our fields was combined the resulting $b/a$ and $c/a$ ratios representing the triaxiality were not constrained as this would require simultaneously fitting the profile in different directions of the sky.

\section{Discussion} \label{sec:discussion}

 In this section, we discuss our findings on the characterization of the overdensity that we identify as the wake in the Galactic halo and the implications for understanding the density profile of the outer halo. % and the mass of the LMC.

\subsection{Comparison with other studies}
The Pisces Overdensity, a large-scale structure in the Milky Way’s outer halo, was initially detected in SDSS Stripe 82 as a stellar overdensity, only surpassed in volume by the Sagittarius Stream \citep{2007AJ....134.2236S}. This region has been mapped using RR Lyrae and blue horizontal branch (BHB) stars, revealing an extended and gradient structure that spans distances from about 40 kpc to 100 kpc \citep{2015ApJ...810..153N, 2009MNRAS.398.1757W, 2018ApJ...862L...1D, 2019MNRAS.488L..47B}. This region was later kinematically associated with the LMC wake by \citep{2019MNRAS.488L..47B}, using radial velocity measurements of 13 BHB stars. 
%$\Delta \rho = 1.6 \pm 0.6$
\citet{2021Natur.592..534C} reported the first detection of the wake, through an all-sky approach to probe the MW halo using K giants and RR Lyrae stars. This was achieved using data from Gaia \citep{2023A&A...674A...1G} and WISE \citep{2010AJ....140.1868W} all-sky surveys. Their findings indicated that the local wake's density contrast exceeded the predictions by \citet{2019ApJ...884...51G} by a factor of \(1.4 \pm 0.2\) and \(2.1 \pm 0.3\), for K giants and RR Lyrae, respectively.  Our study finds an overdensity that is in line with \citet{2019ApJ...884...51G} predictions, consistent between the K giants and nMSTO tracers. The $\Delta \rho$ values reported in \citep{2021Natur.592..534C} differ from those reported in this work, both for the data and the models. This distinction is a combination of two factors: different stellar populations used can trace different densities; and the treatment of the data and simulations done in this work differs in \citet{2021Natur.592..534C}. In particular, the selection function used: whereas in our work we only performed position and photometry based selections (see Section \ref{sec:tracers}), in \citet{2021Natur.592..534C} a proper motion-based selection was applied to both the data and simulations, combined with a re-weighting scheme and masking of regions aligned with the galactic plane ( for a detailed description we refer the reader to methods section in \citet{2021Natur.592..534C} ), these selections will not only affect the observed densities, but most importantly the $\overline{\rho}$, this however does not affect a comparison in terms of the factor by which the prediction is exceeded as that is self-consistent and presented above.
%We note that our methodological procedure for computing the average density of the sky based on radial density profiles is substantially different from the approach used by \citet{2021Natur.592..534C}. 

\citet{amarante24} employed Legacy Survey DR9 \citep{2019AJ....157..168D} photometry to identify a large sample of BHB stars that reach distances of 120 kpc. They report a signature that they identify with the wake through a density contrast $\geq$0.6, measured across a region that is 60 degrees long, 25 degrees wide, and aligned with the LMC’s orbit. They find a radial density profile that follows a double power law but with a break radius and internal/external slopes that vary depending on the direction of the sky (see their figure 7).  This also supports a non-spherical anisotropic stellar halo, in line with our findings presented in Section \ref{sec:radial}. In particular, for the direction of the wake, they fit a double power law to the radial density profile that does not constrain the break radius, which suggests that the break at ~ 70 kpc could also be present in their data. 

\citet{suzuki} analyzed data from the Subaru Hyper Suprime-Cam survey and identified an overdensity in the Pisces region at a heliocentric distance of approximately 60 kpc. Their analysis revealed that the PO extends spatially over several degrees, with a structure indicative of a tidally disrupted stellar system, identified using a matching filtering technique. They measured an increased density of main-sequence stars in this region and found significant alignment with blue straggler stars, suggesting a low metallicity ($[Fe/H] \sim -1.8$) and an age greater than 10 Gyr for the progenitor system. The tidally elongated feature found by \citet{suzuki} is not evident in our data, in particular, a stellar stream sufficiently massive to offset the density as shown in Figure \ref{fig:lmc25_ra} should present a footprint in the radial density profile shown in Figure \ref{fig:radial_b1_25}, as it constitutes an overdensity with consistent galactocentric distance. 

Throughout all these studies, therefore, a detectable overdensity is measured with a statistical significance above $5\sigma$ in a consistent position in the sky. However, \citet{suzuki} find that this overdensity to be consistent with a stream-like feature, which cannot be tested with our survey.

\subsection{Contamination and Substructure}\label{subsec:substructure}

Hierarchical formation models of the Milky Way predict the formation of a spheroidal halo as a result of the assembly of accreted progenitors \citep{1978ApJ...225..357S,1999Natur.402...53H,2005ApJ...635..931B,2006MNRAS.365..747A,2010MNRAS.401.2285G,2012AN....333..460D,2015AJ....150..128R}. Several prominent overdensities have already been detected at distances beyond 50 kpc \citep{2007AJ....134.2236S,2009MNRAS.398.1757W,2019MNRAS.488L..47B}. These overdensities are thought to be remnants of disrupted dwarf galaxies or globular clusters, whose tidal debris can remain coherent in the outer halo for several dynamical times \citep{2017MNRAS.464.2882A,2005ApJ...635..931B,2013MNRAS.436.3602G}. Such substructures could display wake-like footprints and contribute to density variations in the halo. 

For instance, \citet{2023ApJ...956..110C} suggests that at least 20\% and up to 50\% of the Pisces Plume,region that overlaps with Tile 4, is composed of unmixed debris from the LMC and SMC. If this proportion is at the upper limit, accounting for the debris contribution would align both our detection and the findings of \cite{2021Natur.592..534C} with \citet{2019ApJ...884...51G} models. Given the density contrast observed in Figure \ref{fig:lmc25_ra}, reducing it by subtracting potential contamination would, maintain a density contrast consistent with an LMC virial mass at infall of $2.5 \times 10^{11}M_{\odot}$. 

Similarly, \citet{2023ApJ...956..110C} found that debris from the Gaia-Sausage/Enceladus \citep[GSE;][]{2018MNRAS.478..611B, 2018Natur.563...85H} may contribute to the Pisces Overdensity, a region covered by Tiles 2 and 3. This contribution could artificially enhance the observed densities. However, the extent to which this debris affects the densities reported in Figure \ref{fig:lmc25_ra} remains unclear. Notably, Tile 2 exhibits a density consistent with the average halo, whereas Tile 3 shows an enhancement. Further observations are necessary to explore the effect of this substructure.

As shown in Section \ref{sec:ra_dens}, considering the halo models constructed entirely of substructure by \citet{2005ApJ...635..931B} as the null hypothesis for a Z test, our wake detection passes the statistical significance test above 95\%. Therefore, the likelihood that the reported overdensity is the result of substructure is low. However, it is possible that chance alignments of substructure-induced overdensities could reproduce the observed density enhancements, but for this to be the case they must have the same angular size scale as the overdensity we observe, i.e., cover at least 30 degrees.

The most notable substructure that could induce significant contamination beyond 60~kpc in the region of the sky covered by our fields is the Sagittarius stellar stream. As previously mentioned, it could explain the increased density observed within 60~kpc, even though our fields are at higher latitudes than the main track of the Sagittarius stream. Specifically, the two most likely fields to be contaminated are Tile 1 and Tile 2, as they are the closest to the main track of Sagittarius as traced by\citet{2021MNRAS.501.2279V}, with a distance of $\sim 13 \deg$ and $\sim 17 \deg$ respectively. If Tile 1 and Tile 2 were significantly contaminated by the Sagittarius stream, it would imply that the density contrast observed in the line of sight, when using Tile 1 as a comparison, is larger than shown in Figure \ref{fig:radial_b1_25} and that the density differential presented in Figure \ref{fig:lmc25_ra} would also be larger than shown. 

For the distance regime covered by the radial density profiles shown in section \ref{sec:radial}, we need to consider other nearby substructures that could contaminate the radial density profile, mainly the Hercules-Aquila cloud \citep[HAC,][]{hac}. This structure has been found to cover the region between $ 30^{\circ} < l < 60 ^{\circ}$ and $-45^{\circ} < b < 45 ^{\circ}$, at a heliocentric distance between 10 and 20 kpc \citep{hac, hac_comp}. This makes it possible that given the typical distance error of 10\% for nMSTO stars in our data some contamination might come from the HAC, particularly for Tiles 3 and 4, which are the ones closest in projection to the HAC. Given the distance to the HAC, it is very unlikely to have an effect on the density variation across the wake presented in section \ref{sec:ra_dens}, as there is a 40 kpc gap between the stars considered in Figure \ref{fig:lmc25_ra} and the edge of the HAC. 

\subsection{The ambiguity of defining an average halo density}\label{sec:assumptions}

In Section \ref{sec:ra_dens}, we reported the density variation across the sky in terms of $\Delta \rho$, which in the absence of all-sky observational measurements necessitates calculating an average density across the sky within the range of 60–100 kpc by extrapolating radial density profiles. For this analysis, we adopted the best-fit Einasto profile from \citet{2015ApJ...809..144X}, derived from spectroscopically confirmed K giants in the SEGUE survey. This choice facilitates a fair comparison with the predictions by \citet{2019ApJ...884...51G}, which also employ the same radial density profile, making the comparison shown in Figure \ref{fig:lmc25_ra} self-consistent.

Using the radial density profile from \citet{2015ApJ...809..144X} involves two key considerations. First, applying a different profile, such as those proposed by \citet{2008ApJ...680..295B} or \citet{amarante24}, will yield a different average density, thereby rescaling the $\Delta \rho$ values shown in Figure \ref{fig:lmc25_ra}. Indeed, numerous radial density profiles are present in the literature, and as demonstrated in \citet{amarante24}, a line-of-sight-dependent radial density profile may be necessary due to asymmetries in the Milky Way (MW) halo. Therefore, normalizing based on a radial density profile should be considered a comparison against an idealized triaxial halo in the absence of a comprehensive all-sky density measurement up to 100 kpc, which will be available with future photometric and spectroscopic surveys such as LSST \citep{2019ApJ...873..111I}, SDSS-V \citep{2017arXiv171103234K}, DESI \citep{DESI} and WEAVE \citep{weave}. Second, to convert the mass derived from the radial density profile into stellar counts, we assume the MW halo comprises a single, homogeneous stellar population with uniform age and metallicity. In reality, however, the halo contains a range of ages and metallicities, reflecting the various substructures that constitute it.

In the same way that the value of $\overline{\rho}$ can change, the results from the Z-test performed in section \ref{sec:ra_dens} can change, in the worst case scenario of the entire sky having an average density of the Pisces Overdensity region, we can still expect the peak density to be $1 \sigma$ beyond the density variations found in \citet{2005ApJ...635..931B}.

\section{Summary and conclusions} \label{sec:conclusions}

We obtained wide-field deep optical and near-infrared imaging with the DECam and VIRCAM instruments, respectively, in four fields covering the region of the Southern sky expected to show a stellar density enhancement in the Galactic halo due to the presence of the dynamical friction wake caused by the passage of the LMC \citep[e.g.,][]{2019ApJ...884...51G}.  These data reach significantly fainter levels than previous surveys that cover this region, allowing us to map distant regions of the MW halo with good number statistics (provided by faint tracers), and using two different stellar tracers.  Taking advantage of $(g-i)$, $(i-K_{\rm S})$ color-color selections, an efficient and reliable star/background galaxy separation is achieved in regions of the CMD previously prohibited by the high levels of confusion (see Section \ref{subsec:stargal} and Figure \ref{fig:stars_sigma}).  We obtain samples of 102 K giants and 309 nMSTO stars spanning Galactocentric distances between $60-100$ kpc.  Our analysis of  these datasets has revealed the following:

\begin{itemize}

 \item The radial stellar halo density profile in the region we examine is consistent with the outer slope of the double power-law fit by \citet{2008ApJ...680..295B}, which was derived using MSTO stars from SDSS imaging. However, this agreement holds only up to a distance of $60-70$ kpc from the Galactic center (Figure \ref{fig:radial_b1_25}). Beyond this range, the density profile becomes steeper across all our fields, aligning more closely with the Einasto-like profile identified by \citet{2015ApJ...809..144X} using K giants from the SEGUE survey. It is worth noting that both \citet{2008ApJ...680..295B} and \citet{2015ApJ...809..144X} focused on the Galactic halo as observed from the Northern hemisphere, where the amplitude of the perturbations induced by the LMC are expected to be lower. When we combine the nMSTO stars from all sightlines, this transition becomes more pronounced, revealing a clear break radius at approximately $70$ kpc, deep in the distant halo (Figure \ref{fig:fit}).

 \item The integrated number of halo stars with Galactocentric distances between $60-100$ kpc steadily increases across the sky with increasing Right Ascension (Figure \ref{fig:lmc25_ra}, going from our Tile 1 to Tile 4), which corresponds to moving across the Southern sky into the expected current location of the LMC wake according to numerical simulations.  Both stellar tracers used, K giants and nMSTO stars, show this same behavior, in qualitative and quantitative terms.

 \item  The maximum density enhancement observed in our data, relative to a smooth local halo, occurs in the easternmost of our fields, which prevents us from determining whether we have reached the actual peak of the overdensity associated with the LMC wake. The density contrast at this maximum is $\Delta \rho = 1.6 \pm 0.6$, corresponding to roughly three times the average density of a smooth local halo, as estimated in Section \ref{sec:ra_dens}, with the caveats and assumptions of this measurement presented in section \ref{sec:assumptions}. This overdensity is as pronounced as that found in the most massive LMC model ($2.5 \times 10^{11}M_{\odot}$) and consistent at the $1\sigma$ level with the value reported by \cite{2021Natur.592..534C} from their all-sky analysis using K giants.

 \item As a function of Galactocentric distance, the radial halo density profiles, normalized to that of our field farthest from the wake (Figure \ref{fig:radial_b1_25}), show a density contrast/enhancement beyond 70 kpc that steadily increases as we move across the sky towards the expected current location of the LMC wake, as predicted by simulations. This result is consistent with the findings from the integrated star counts.

 \item At closer distances (20-60 kpc), where models of the LMC infall do not predict a significant contribution from the wake, we still observe an enhancement in the radial density profiles relative to our reference field (the field farthest from the wake's location in the sky). This enhancement is small near the reference field but increases rapidly towards the east, in the direction of the projected wake location. Since current models do not expect the LMC wake to influence the inner regions of the stellar halo as covered by our data, these excesses may suggest the presence of substructure from other origins or a highly anisotropic, non-spherical halo in these regions of the sky, with the most likely culprit being the HAC-S overdensity \citep{hac}, even though the the HAC-S lies between 10 - 20 kpc.

\end{itemize}

Most of the findings summarized above regarding the distant Galactic stellar halo in the direction of our fields align with predictions from current numerical simulations of the LMC's infall towards the Milky Way \citep{2019ApJ...884...51G}. This provides strong evidence that the overdensity observed in our survey corresponds to the stellar wake trailing the LMC during its first passage through the halo. Additionally, the observed density enhancements could potentially be influenced by other forms of halo substructure in the region we probed. The presence of some contamination is further supported by the identification of stellar debris from the LMC and SMC in the Pisces Plume region \citep{2023ApJ...956..110C}, debris from GSE in the Pisces Overdensity \citep{2023ApJ...951...26C}, as well as by the excess we observe at distances closer than the current location of the Clouds.

This work has explored the distant MW stellar halo down to photometric depths previously unprobed in the Southern hemisphere through wide-field observations. Future studies with broader area coverage will be essential for further characterizing the LMC wake and disentangling potential contamination from other substructures in this region. In particular, more detailed investigations of the LMC wake and halo substructure will be enabled by upcoming surveys such as LSST \citep{2019ApJ...873..111I}, which will provide multiband photometry and proper motion data for much of the Southern hemisphere. When combined with radial velocities and chemical abundances from spectroscopic surveys like SDSS-V \citep{2017arXiv171103234K}, DESI \citep{DESI} and WEAVE \citep{weave}, these data will offer a comprehensive, multidimensional view of the Milky Way's halo. This will improve our understanding of the Galaxy's assembly, the mass and orbital elements of its more massive satellites, and ultimately allow the use of their interactions as a test for dark matter models and gravity itself.

%facilitating the measurement of other predicted effects of the LMC perturbation, such as the reflex motion of the halo and the DF wake kinematic response, as well as resolving issues with contamination from substructures or debris from the Clouds.
%Thanks to the depth and quality of our data we reach $K_s \approx 24.5$, which translates to $\approx 100$ kpc distance limit for nMSTO stars. 

%Our novel star/galaxy separation procedure allows us to utilize faint abundant tracers in regions of the CMD previously prohibited by the high levels of contamination by non-resolved point source like background galaxies. 

%We calculate the galactocentric distances based on a flat prior photometric distance modulus of nMSTO and giant stars. 

%We binned nMSTO stars by galactocentric distance and calculated the stellar number density distribution along the 4 lines of sight,

\section*{Acknowledgements}
We thank Kathryn Johnston, Manuela Zoccali, Javier Alonso, Roberto Muñoz, Rogelio Albarracin, Aquiles den Braber, Alvaro Valenzuela, Eitan Dvorquez and Julio Olivares for the insightful discussions.  MC and JC acknowledge support from the Agencia Nacional de Investigación y Desarrollo (ANID) via Proyecto Fondecyt Regular 1191366 and 1231345, and by ANID BASAL project FB210003. CN acknowledges support by the Centre National d'\'etudes Spatiales (CNES). G.B. is supported by NSF CAREER AST1941096. FAG. acknowledge funding from the Max Planck Society through a “PartnerGroup” grant. FAG acknowledges support from ANID FONDECYT Regular 1211370, the ANID Basal Project FB210003 and the HORIZON-MSCA-2021-SE-01 Research and innovation programme under the Marie Sklodowska-Curie grant agreement number 101086388.  The Geryon cluster at the Centro de Astro-Ingenieria UC was extensively used for the calculations performed in this work. ANID BASAL project FB21000, BASAL CATA PFB-06, the Anillo ACT-86, FONDEQUIP AIC-57, and QUIMAL 130008 provided funding for several improvements to the Geryon cluster.

\appendix
\label{subsec:apendix}

\section{Coefficients for the star/galaxy separation}

Polynomial best-fit found for the star/galaxy separation procedure described in Section~\ref{subsec:stargal}.
\begin{equation}
\label{eq:poly_stars}
\begin{split}
i - Ks = &-8.90365315 \times 10^{-4} (g - i)^6 \\
&+ 5.40619658 \times 10^{-3} (g - i)^5 \\
&+ 6.25122688 \times 10^{-3} (g - i)^4 \\
&- 4.85818006 \times 10^{-2} (g - i)^3 \\
&- 0.116253605 (g - i)^2 \\
&+ 1.04653713 (g - i) \\
&- 0.961426796
\end{split}
\end{equation}

\section{Fitting procedure} \label{sec:mcmc}

We performed the fit using a Markov Chain Monte Carlo (MCMC) sampling method to estimate the parameters of the triaxial broken power-law density model:

\begin{equation}\label{eq:bpl}
\rm \rho_\star\left(r_q \right )= \left\{\begin{matrix} \rm
\rho_0~r_q^{-\alpha_{in}}, & \rm r_q\leq r_{break} \\ \rm \nu_0\times
r_{break}^{(\alpha_{out}-\alpha_{in})}\times r_q^{-\alpha_{out}},&\rm
r_q>r_{break} \end{matrix}\right.
\end{equation}

The log-likelihood function used is \(\mathcal{L} \propto -0.5 \sum \left(\frac{\rho_{\text{model}} - \rho_{\text{data}}}{\sigma_{\text{data}}}\right)^2 + \log(2 \pi \sigma_{\text{data}}^2)\), where \(\sigma_{\text{data}}\) represents the uncertainty in the density counts derived from distance uncertainties through Monte Carlo resampling. We assigned Gaussian priors for the variables \(A\), \(b/a\), \(c/a\), \(\alpha_{\text{in}}\), \(\alpha_{\text{out}}\), and \(r_{\text{break}}\), centered on empirically derived values—namely, 1, 1, 1, 3.08, 7.5, and 6.8 kpc, respectively. These priors ensure physically motivated constraints, while still allowing for variations supported by our data. We used a binned dataset for the nMSTO stars, with uncertainties on each bin derived from the propagated distance errors. The MCMC sampling was performed using the `emcee` library \citep{2013PASP..125..306F}, which implements an affine-invariant ensemble sampler. A total of 32 walkers were run over 10,000 steps, with the first 1,000 steps discarded as burn-in. We provide a corner plot showing the posterior distribution of the parameters to visualize correlations between them. All best-fit parameters were extracted from the MCMC chains by marginalizing over the posterior distributions.

\begin{figure}
    \centering
    \includegraphics[width=0.7\linewidth]{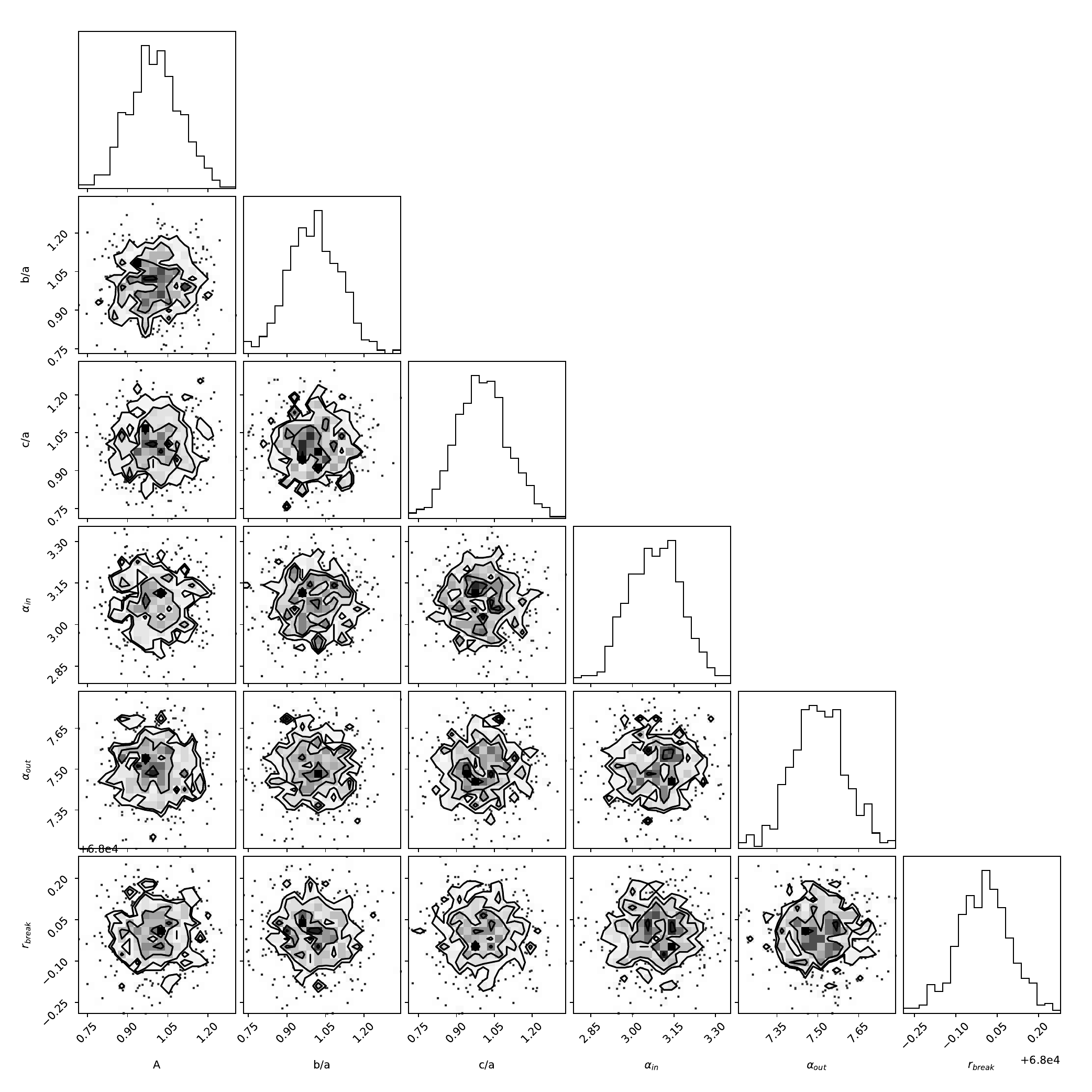}
    \caption{Corner plot showing the posterior distributions of the parameters from the MCMC fit to the triaxial broken power-law density model for the nMSTO stars in our 4 observed fields. The plot displays marginalized one-dimensional histograms for each parameter and two-dimensional contours for parameter pairs. The parameters include \(A\), \(b/a\), \(c/a\), \(\alpha_{\text{in}}\), \(\alpha_{\text{out}}\), and \(r_{\text{break}}\)  The contours represent $1\sigma$, $2\sigma$, and $3\sigma$ confidence levels, derived after discarding the first 1,000 burn-in steps from the MCMC chains.}
    \label{fig:corner_plot}
\end{figure}

%\clearpage
\bibliography{sample631}{}
\bibliographystyle{aasjournal}

\end{document}